\documentclass[12pt]{article}

\usepackage{amsmath,amsfonts,amsbsy,enumerate,array}
\usepackage{graphicx}
\usepackage[usenames]{color}
\usepackage[english]{babel}
\usepackage{fancybox}
\usepackage{chngpage} 

\newcommand{\captionfonts}{\small}

\makeatletter  
\long\def\@makecaption#1#2{%
  \vskip\abovecaptionskip
  \sbox\@tempboxa{{\captionfonts #1: #2}}%
 \ifdim \wd\@tempboxa >\hsize
    {\captionfonts #1: #2\par}
  \else
    \hbox to\hsize{\hfil\box\@tempboxa\hfil}%
  \fi
  \vskip\belowcaptionskip}
\makeatother   

\usepackage{mciteplus}

\usepackage[colorlinks=true,      linkcolor=blue,      urlcolor=blue,      filecolor=blue,      citecolor=blue,
      pdfstartview=FitH,     pdfpagemode=UseNone,      bookmarksopen=true]{hyperref}  
      
\usepackage[all]{hypcap}     

\begin{document}

\numberwithin{equation}{section}


\mathchardef\mhyphen="2D


\newcommand{\be}{\begin{equation}}
\newcommand{\ee}{\end{equation}}
\newcommand{\bea}{\begin{eqnarray}\displaystyle}
\newcommand{\eea}{\end{eqnarray}}
\newcommand{\nnm}{\nonumber}
\newcommand{\nn}{\nonumber}
\newcommand{\newotimes}{}  
\newcommand{\utv}{|0_R^{-}\rangle^{(1)}\newotimes |0_R^{-}\rangle^{(2)}}
\newcommand{\utvb}{|\bar 0_R^{-}\rangle^{(1)}\newotimes |\bar 0_R^{-}\rangle^{(2)}}

\def\eq#1{(\ref{#1})}
\newcommand{\secn}[1]{Section~\ref{#1}}

\newcommand{\tbl}[1]{Table~\ref{#1}}
\newcommand{\fig}{Fig.~\ref}

\def\bet{\begin{tabular}}
\def\eet{\end{tabular}}
\def\bs{\begin{split}}
\def\es{\end{split}}


\def\a{\alpha}  \def\b{\beta}   \def\c{\chi}
\def\g{\gamma}  \def\G{\Gamma}  \def\e{\epsilon}
\def\vep{\varepsilon}   \def\tvep{\widetilde{\varepsilon}}
\def\f{\phi}    \def\F{\Phi}  \def\fb{{\ov \phi}}
\def\vf{\varphi}  \def\m{\mu}  \def\mub{\ov \mu}
\def\n{\nu}  \def\nub{\ov \nu}  \def\o{\omega}
\def\O{\Omega}  \def\r{\rho}  \def\k{\kappa}
\def\kab{\ov \kappa}  \def\s{\sigma}
\def\t{\tau}  \def\th{\theta}  \def\sb{\ov\sigma}  \def\S{\Sigma}
\def\l{\lambda}  \def\L{\Lambda}  \def\p{\psi}

\newcommand{\gt}{\tilde{\gamma}}


\def\cA{{\cal A}} \def\cB{{\cal B}} \def\cC{{\cal C}}
\def\cD{{\cal D}} \def\cE{{\cal E}} \def\cF{{\cal F}}
\def\cG{{\cal G}} \def\cH{{\cal H}} \def\cI{{\cal I}}
\def\cJ{{\cal J}} \def\cK{{\cal K}} \def\cL{{\cal L}}
\def\cM{{\cal M}} \def\cN{{\cal N}} \def\cO{{\cal O}}
\def\cP{{\cal P}} \def\cQ{{\cal Q}} \def\cR{{\cal R}}
\def\cS{{\cal S}} \def\cT{{\cal T}} \def\cU{{\cal U}}
\def\cV{{\cal V}} \def\cW{{\cal W}} \def\cX{{\cal X}}
\def\cY{{\cal Y}} \def\cZ{{\cal Z}}

\def\mC{\mathbb{C}} \def\mP{\mathbb{P}}
\def\mR{\mathbb{R}} \def\mZ{\mathbb{Z}}
\def\mT{\mathbb{T}} \def\mN{\mathbb{N}}
\def\mH{\mathbb{H}} \def\mX{\mathbb{X}}

\def\CP{\mathbb{CP}}
\def\RP{\mathbb{RP}}
\def\Z{\mathbb{Z}}
\def\N{\mathbb{N}}
\def\H{\mathbb{H}}

\newcommand{\rmd}{\mathrm{d}}
\newcommand{\rmx}{\mathrm{x}}

\def\tA{ {\widetilde A} }

\def\one{{\hbox{\kern+.5mm 1\kern-.8mm l}}}
\def\zero{{\hbox{0\kern-1.5mm 0}}}


\newcommand{\bra}[1]{{\langle {#1} |\,}}
\newcommand{\ket}[1]{{\,| {#1} \rangle}}
\newcommand{\braket}[2]{\ensuremath{\langle #1 | #2 \rangle}}
\newcommand{\Braket}[2]{\ensuremath{\langle\, #1 \,|\, #2 \,\rangle}}
\newcommand{\norm}[1]{\ensuremath{\left\| #1 \right\|}}
\def\corr#1{\left\langle \, #1 \, \right\rangle}
\def\vac{|0\rangle}


\def\d{ \partial }
\def\zb{{\bar z}}

\newcommand{\sq}{\square}
\newcommand{\IP}[2]{\ensuremath{\langle #1 , #2 \rangle}}    

\newcommand{\floor}[1]{\left\lfloor #1 \right\rfloor}
\newcommand{\ceil}[1]{\left\lceil #1 \right\rceil}

\newcommand{\dbyd}[1]{\ensuremath{ \frac{\d}{\d {#1}}}}
\newcommand{\ddbyd}[1]{\ensuremath{ \frac{\d^2}{\d {#1}^2}}}

\newcommand{\Zd}{\ensuremath{ Z^{\dagger}}}
\newcommand{\Xd}{\ensuremath{ X^{\dagger}}}
\newcommand{\Ad}{\ensuremath{ A^{\dagger}}}
\newcommand{\Bd}{\ensuremath{ B^{\dagger}}}
\newcommand{\Ud}{\ensuremath{ U^{\dagger}}}
\newcommand{\Td}{\ensuremath{ T^{\dagger}}}

\newcommand{\T}[3]{\ensuremath{ #1{}^{#2}_{\phantom{#2} \! #3}}}		

\newcommand{\tr}{\operatorname{tr}}
\newcommand{\sech}{\operatorname{sech}}
\newcommand{\Spin}{\operatorname{Spin}}
\newcommand{\Sym}{\operatorname{Sym}}
\newcommand{\Com}{\operatorname{Com}}
\def\adj{\textrm{adj}}
\def\id{\textrm{id}}

\def\ha{\frac{1}{2}}
\def\tha{\tfrac{1}{2}}
\def\wt{\widetilde}
\def\ra{\rangle}
\def\la{\langle}

\def\pb{\ov\psi}
\def\pt{\widetilde{\psi}}
\def\at{\widetilde{\a}}
\def\cb{\ov\chi}
\def\d{\partial}
\def\db{\bar\partial}
\def\delb{\bar\partial}
\def\dbar{\ov\partial}
\def\dag{\dagger}
\def\dalpha{{\dot\alpha}}
\def\dbeta{{\dot\beta}}
\def\dgamma{{\dot\gamma}}
\def\ddelta{{\dot\delta}}
\def\ad{{\dot\alpha}}
\def\bd{{\dot\beta}}
\def\dg{{\dot\gamma}}
\def\dd{{\dot\delta}}
\def\th{\theta}
\def\Th{\Theta}
\def\eb{{\ov \epsilon}}
\def\gb{{\ov \gamma}}
\def\wb{{\ov w}}
\def\Wb{{\ov W}}
\def\D{\Delta}
\def\DD{\Delta^\dag}
\def\Db{\ov D}

\def\ov{\overline}
\def\Slash{\, / \! \! \! \!}
\def\dslash{\partial\!\!\!/}
\def\Dslash{D\!\!\!\!/\,\,}
\def\fslash#1{\slash\!\!\!#1}
\def\Fslash#1{\slash\!\!\!\!#1}

\def\del{\partial}
\def\delb{\bar\partial}
\newcommand{\ex}[1]{{\rm e}^{#1}}
\def\ii{{i}}

\newcommand{\vs}[1]{\vspace{#1 mm}}

\newcommand{\ve}{{\vec{\e}}}
\newcommand{\shalf}{\frac{1}{2}}

\newcommand{\lb}{\rangle}
\newcommand{\al}{\ensuremath{\alpha'}}
\newcommand{\ap}{\ensuremath{\alpha'}}

\newcommand{\bean}{\begin{eqnarray}}
\newcommand{\eean}{\end{eqnarray}}
\newcommand{\ft}[2]{{\textstyle {\frac{#1}{#2}} }}

\newcommand{\hsp}{\hspace{0.5cm}}
\def\half{{\textstyle{1\over2}}}
\let\ci=\cite \let\re=\ref
\let\se=\section \let\sse=\subsection \let\ssse=\subsubsection

\newcommand{\dpb}{D$p$-brane}
\newcommand{\dpbs}{D$p$-branes}

\def\gh{{\rm gh}}
\def\sgh{{\rm sgh}}
\def\NS{{\rm NS}}
\def\R{{\rm R}}
\def\Qp{Q_{\rm P}}
\def\QP{Q_{\rm P}}

\newcommand\dott[2]{#1 \! \cdot \! #2}

\def\eo{\overline{e}}


\def\p{\partial}
\def\h{{1\over 2}}

\def\d{\partial}
\def\la{\lambda}
\def\eps{\epsilon}
\def\bb{\bigskip}
\def\tg{\widetilde\gamma}
\newcommand{\dm}{\begin{displaymath}}
\newcommand{\edm}{\end{displaymath}}
\renewcommand{\b}{\widetilde{B}}
\newcommand{\gm}{\Gamma}
\newcommand{\ac}[2]{\ensuremath{\{ #1, #2 \}}}
\renewcommand{\ell}{l}
\newcommand{\z}{\ell}
\def\bb{$\bullet$}
\def\Qbar{{\bar Q}_1}
\def\QPbar{{\bar Q}_p}

\def\q{\quad}

\def\bn{B_\circ}

\let\a=\alpha \let\b=\beta \let\g=\gamma
\let\e=\epsilon
\let\c=\chi \let\th=\theta  \let\k=\kappa
\let\l=\lambda \let\m=\mu \let\n=\nu \let\x=\xi \let\r=\rho

\let\s=\sigma

\let\vp=\varphi \let\vep=\varepsilon
\let\w=\omega  \let\G=\Gamma \let\D=\Delta \let\Th=\Theta \let\P=\Pi \let\S=\Sigma

\def\h{{1\over 2}}

\def\r{\rightarrow}
\def\Ri{\Rightarrow}

\def\nn{\nonumber\\}
\let\bm=\bibitem
\def\Kt{{\widetilde K}}
\def\b{\vspace{3mm}}
\def\t{\tilde}
\let\p=\partial

\def\sqi{{1\over \sqrt{2}}}
\newcommand\com[2]{[#1,\,#2]}


\newcommand{\spectral}{Schwimmer:1986mf}

\newcommand{\sv}{Strominger:1996sh}
\newcommand{\adscft}{Maldacena:1997re,*Witten:1998qj,*Gubser:1998bc}

\newcommand{\cvet}{Cvetic:1996xz,*Cvetic:1997uw}
\newcommand{\mm}{Balasubramanian:2000rt,*Maldacena:2000dr}
\newcommand{\MM}{Balasubramanian:2000rt,*Maldacena:2000dr}

\newcommand{\lmone}{Lunin:2001ew}
\newcommand{\lmtwo}{Lunin:2001pw}
\newcommand{\lmfour}{Lunin:2001jy}
\newcommand{\lmRot}{Lunin:2001fv}
\newcommand{\lmAdS}{Lunin:2001jy}
\newcommand{\lmm}{Lunin:2002iz}

\newcommand{\GMR}{Gutowski:2003rg}
\newcommand{\CarMcCon}{Cariglia:2004wu}

\newcommand{\lunin}{Lunin:2004uu}
\newcommand{\gmsone}{Giusto:2004id}
\newcommand{\gmstwo}{Giusto:2004ip}

\newcommand{\bw}{Bena:2004de}
\newcommand{\fuzz}{
Mathur:2005zp,*Bena:2007kg,*Skenderis:2008qn,*Balasubramanian:2008da,*Chowdhury:2010ct,Mathur:2012zp,*Bena:2013dka}
\newcommand{\kst}{Kanitscheider:2007wq}
\newcommand{\cern}{Mathur:2009hf}

\newcommand{\mtone}{Mathur:2011gz}
\newcommand{\mttwo}{Mathur:2012tj}
\newcommand{\lmt}{Lunin:2012gp}
\newcommand{\glmt}{Giusto:2012yz}

\newcommand{\acmtwo}{Avery:2010er}
\newcommand{\acmthree}{Avery:2010hs}

\newcommand{\chmtone}{Carson:2014yxa}
\newcommand{\cmt}{Carson:2014xwa}

\newcommand{\orbifoldrefs}{Arutyunov:1997gt,*Arutyunov:1997gi,*deBoer:1998ip,*Dijkgraaf:1998gf,
*Seiberg:1999xz,*Larsen:1999uk,*David:1999zb,*Jevicki:1998bm,*Lunin:2002fw,*Gomis:2002qi,*Gava:2002xb}

\newcommand{\orbifoldtwo}{Pakman:2009zz,*Pakman:2009ab,*Pakman:2009mi,*Keller:2011xi,*Cardona:2014gqa}

\newcommand{\twocharge}{Balasubramanian:2000rt,*Maldacena:2000dr,Lunin:2001jy,*Lunin:2002bj,Skenderis:2006ah,*Kanitscheider:2006zf,*Kanitscheider:2007wq}

\newcommand{\fuzzrecentone}{Bena:2011uw,*Giusto:2011fy,*Gibbons:2013tqa}

\newcommand{\fuzzrecenttwo}{Mathur:2013nja,*Bena:2013ora}

\newcommand{\fuzzrecentthree}{Bena:2014qxa,*Chen:2014loa,*Martinec:2014gka}


\begin{flushright}
IPhT-T14/154
\end{flushright}

\vspace{16mm}

 \begin{center}
{\LARGE Effect of the deformation operator in the D1D5 CFT}
\\
\vspace{15mm}
{\bf  Zaq Carson$^1$, Shaun Hampton$^1$, Samir D. Mathur$^1$ and David Turton$^2$ 
\\}
\vspace{10mm}
$^1\,$Department of Physics,\\ The Ohio State University,\\ 
191 West Woodruff Avenue,\\
Columbus, OH 43210, USA\\ 
\vspace{5mm}

$^2\,$Institut de Physique Th\' eorique, \\
CEA Saclay, CNRS URA 2306 \\
91191 Gif-sur-Yvette, France

\vspace{8mm}

carson.231@osu.edu, hampton.197@osu.edu, \\ 
mathur.16@osu.edu, david.turton@cea.fr

\end{center}

\vspace{4mm}

\thispagestyle{empty}
\begin{abstract}

\vspace{3mm}

The D1D5 CFT gives a holographic dual description of a near-extremal black hole in string theory. The interaction in this theory is given by a marginal deformation operator, which is composed of supercharges acting on a twist operator. The twist operator links together different copies of a free CFT. We study the effect of this deformation operator when it links together CFT copies with winding numbers $M$ and $N$ to produce a copy with winding $M+N$, populated with excitations of a particular form. We compute the effect of the deformation operator in the full supersymmetric theory, firstly on a Ramond-Ramond ground state and secondly on states with an initial bosonic or fermionic excitation. Our results generalize recent work which studied only the bosonic sector of the CFT. Our findings are a step towards understanding thermalization in the D1D5 CFT, which is related to black hole formation and evaporation in the bulk.

\end{abstract}
\newpage

\baselineskip=15pt
\parskip=3pt
\setcounter{footnote}{0}

\section{Introduction}\label{intr}

String theory has successfully described many aspects of the quantum physics of black holes. In string theory we have an explicit microscopic description of a near-extremal black hole, given by the D1D5 system -- a bound state of $N_1$ D1 branes and $N_5$ D5 branes. This bound state and its excitations reproduce the entropy of black holes in 4+1 noncompact dimensions \cite{Strominger:1996sh}, and provides one of the original examples of AdS/CFT duality~\cite{Maldacena:1997re,*Witten:1998qj,*Gubser:1998bc}.

The gauge theory on the D1D5 bound state flows in the infrared to a (4,4) SCFT. It is conjectured that there is a point in the moduli space where this SCFT is a symmetric product orbifold theory, consisting of $N_1N_5$ symmetrized copies of a free $c=6$ SCFT \cite{\orbifoldrefs}.

At this `orbifold point' in moduli space, the gravity description is strongly coupled. 
%
Despite this fact, much has been achieved by studying the orbifold CFT. There is a well-studied AdS/CFT dictionary for two-charge BPS states \cite{\twocharge}, as well as three-charge BPS states arising from spectral flow \cite{Lunin:2004uu,*Giusto:2004id,*Giusto:2004ip}. More recently, gravity solutions corresponding to the symmetry algebra generators of the CFT have been constructed \cite{\mtone,*\mttwo,*\lmt,Giusto:2013bda}, and CFT duals of the most general two-center D1D5P gravity solutions \cite{Jejjala:2005yu,Bena:2005va,*Berglund:2005vb} have been identified \cite{Giusto:2012yz}. The orbifold CFT has even been shown to describe aspects of non-BPS states, including reproducing the radiation rate from non-BPS gravity solutions \cite{Jejjala:2005yu,Cardoso:2005gj,Chowdhury:2007jx,*Chowdhury:2008bd,*Chowdhury:2008uj}. 

There are of course physical processes which are not well described by the orbifold CFT. In particular, it is interesting to consider the process of black hole formation in the bulk, which is dual to a process of thermalization in the CFT\footnote{Here we are using the term `thermalization' in a somewhat broad sense; it has been argued~\cite{Dimitrakopoulos:2014ada} that in global AdS, below the Hawking-Page transition, black hole formation should correspond to prethermalization~\cite{Berges:2004ce}, and the subsequent evaporation should correspond to actual thermalization.}. To study thermalization, we must use the deformed theory. The operator content of the undeformed theory includes twist operators; a twist operator $\sigma_M$ takes $M$ copies of the $c=6$ CFT and joins them into a single copy living on a circle which is $M$ times longer. We denote any such linked set of copies a `component string'. The marginal deformation is composed of a twist operator $\sigma_2^{++}$ dressed with left- and right-moving supercharges: $\hat O\sim G_{-\h}\bar{G}_{-\h} \sigma^{++}_2$ (see e.g. \cite{David:2002wn}).

\begin{figure}[t]
\begin{center}
\includegraphics[scale=.5]{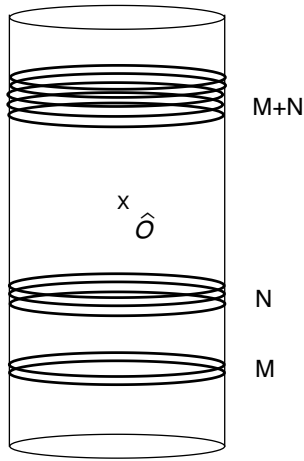}
\caption{The process of interest, sketched on the cylinder. 
The state before the insertion of the deformation operator has component strings with windings $M,N$. 
The deformation operator $\hat O$ links these into a single component string of winding $M+N$.}
\label{fone}
\end{center}
\end{figure}

We are interested in the process in which the deformation operator joins component strings of length $ M $ and $ N $ to make a component string of length $ M+N $\footnote{The twist operator $\sigma^{++}_2$ may also act on two strands of the same component string, whereupon it will divide the component string into two parts. The computations for this case can be carried out in a similar manner to the calculations in the present paper.}. This process is depicted in Fig.\;\ref{fone}. For $M=N=1$, the effect of the deformation operator has has been computed, firstly in the case where the operator acts on the vacuum state of the two initial component strings \cite{\acmtwo}, and also in the case where the operator acts on excited states \cite{\acmthree}. Recently, focusing on bosonic fields only and using the bosonic part of the twist operator $\sigma_2$, this process was studied for general $M$, $N$, both using exact CFT techniques \cite{\chmtone} and also using an alternative method involving  Bogoliubov coefficients, which obtained the effect of the twist in a `continuum limit' approximation \cite{\cmt}. We shall discuss this approximation in more detail later.

In this paper we compute the effect of the deformation operator in the above process, in the full supersymmetric theory. As well as analyzing the effect of the deformation operator on the fermionic fields, we compute the overall prefactor of the final state, which was not computed in the bosonic case of \cite{\chmtone}.

In more detail, we carry out the following calculations:

(a) Firstly, we begin with a given Ramond-Ramond (R-R) ground state $|0^{--}_R\rangle^{(1)} |0^{--}_R\rangle^{(2)}$ on the initial component strings depicted in Fig.\;\ref{fone} (we will define all notation in due course). We apply the deformation operator at the point $w_0$. The deformation operator consists of supercharges acting on a twist operator $\sigma^{++}_{2}$; we first apply the twist operator. This generates a component string with winding $M+N$. 
Following the arguments in \cite{\acmtwo}, we make the ansatz that the state produced by the twist $\s^{++}_2$ on the final component string has the schematic form of an exponential acting on the R-R ground state $|0^{--}_R\rangle$:
\bea
\ket{\boldsymbol{\chi}} &\equiv&
\sigma_{2}^{++}(w_{0})
|0^{--}_R\rangle^{(1)} |0^{--}_R\rangle^{(2)} \cr
&=&
 C_{MN}  
\left[ \exp\left({\sum_{k,l}\gamma^B_{kl}\alpha_{-k}\alpha_{-l}
+\sum_{k,l}\gamma^F_{kl}d_{-k}d_{-l}}\right) 
\times [\mathrm{antihol.}] \right]
\ket{0_{R}^{--}}
\eea
where $\alpha_k$ and $d_k$ are the oscillator modes of a free boson and fermion respectively. 
For the case where the initial component strings had windings $M=N=1$, the coefficients $\gamma^B_{kl}$, $\gamma^F_{kl}$ and $C_{11}=1$ were computed  in \cite{\acmtwo}, and the exponential ansatz was verified up to fourth order in the bosonic oscillators. 
For general $M,N$, the bosonic coefficients $\gamma^B_{kl}$ were calculated in \cite{\chmtone}.
In the present paper we will compute the fermionic coefficients $\gamma^F_{kl}$ and the overall prefactor $C_{MN}$ for arbitrary $M,N$. The calculation of $C_{MN}$ is quite nontrivial and involves the methods of \cite{\lmone,\lmtwo}.
We then apply the supercharge to obtain the full effect of the deformation operator on the R-R ground state $|0^{--}_R\rangle^{(1)} |0^{--}_R\rangle^{(2)}$. 

\b

(b) Secondly, we begin with an initial excitation $\alpha^{(i)}_{-m}$ or $d^{(i)}_{-m}$ on one of the component strings before the twist; here the index $i=1,2$ labels the component string which we excite. After the action of the twist, each such excitation is converted into a linear combination of excitations above the state $|\boldsymbol{\chi}\rangle$,
\bea
\sigma_{2}^{++}(w_{0})\alpha^{(i)}_{-m} |0^{--}_R\rangle^{(1)} |0^{--}_R\rangle^{(2)} ~&=&~\sum\limits_k f^{B(i)}_{mk}\,  \alpha_{-k}|\boldsymbol{\chi}\rangle \,, \qquad i=1,2 \cr
\sigma_{2}^{++}(w_{0})d^{(i)}_{-m} |0^{--}_R\rangle^{(1)} |0^{--}_R\rangle^{(2)} ~&=&~\sum\limits_k f^{F(i)}_{mk}\,  d^{(i)}_{-k}|\boldsymbol{\chi}\rangle \,, \qquad i=1,2
\eea
on the final component string of length $M+N$. 
(Due to the SU(2) indices on the fermions, which are suppressed in the above schematic expressions, there are two fermionic quantities $f^{F\pm(i)}$ which we will define in due course).
For $M=N=1$, the coefficients $f^{B(i)}_{mk}$ and $f^{F(i)}_{mk}$ were found in \cite{\acmthree}. 
For general $M,N$, the coefficients $f^{B(i)}_{mk}$ were calculated in \cite{\chmtone}.
In the present paper, for general $M,N$, we compute the coefficients $f^{F(i)}_{mk}$ and apply the supercharge in both bosonic and fermionic cases, thus obtaining the full effect of the deformation operator on an initial bosonic or fermionic excitation. 

\b

(c) The regime of parameters of interest for  black hole physics is  $N_1N_5\gg 1$. In this regime, component strings typically have windings $M\gg1$, which is the main physical reason for studying the present general $ M$, $ N $ problem.
In this limit, the excitations on such component strings typically have wavelengths much shorter than the length of the component string. 
Thus we are interested in taking a `continuum limit' in which the excitations $\alpha^{(i)}_{-m}$, $d^{(i)}_{-m}$ have $m\gg 1$; such excitations are not sensitive to the infra-red cutoff scale set by the length of the component string. We find the continuum limit approximations to the expressions for $\gamma^F, f^{F(i)}$.

The deformation operator has recently been studied in various other works. Intertwining relations for operators before and after a twist insertion were derived in \cite{Avery:2010vk}. The effect of the twist on entanglement entropy was studied in \cite{Asplund:2011cq}. Twist-nontwist correlators were studied in \cite{Burrington:2012yn}, and operator mixing was computed in \cite{Burrington:2012yq}. 
For other related work, see \cite{\orbifoldtwo}. This line of enquiry complements the fuzzball program \cite{\fuzz}; for recent work in this area, see e.g.~\cite{\fuzzrecentone,\fuzzrecenttwo,\fuzzrecentthree}.

This paper is organized as follows. 
In Section \ref{sectiontwo} we introduce the D1D5 CFT. 
In Section \ref{sec:intrcomp} we introduce the computation. 
Section \ref{computation} describes the method we use. 
In Section \ref{sec:gamma} we compute the effect of the deformation operator on the R-R ground state $|0^{--}_R\rangle^{(1)} |0^{--}_R\rangle^{(2)}$.
In Section \ref{sec:f} we compute the effect of the deformation operator on states with initial excitations. 
In Section \ref{sec:cont} we obtain the continuum limit approximations to the quantities $\gamma^F, f^{F(i)}$. 
In Section \ref{sec:disc} we discuss our results.

\section{The D1D5 CFT} \label{sectiontwo}
Here we review some of the properties of the D1D5 CFT.  First, consider type IIB string theory compactified as:
\be
M_{9,1}\rightarrow M_{4,1}\times S^1\times T^4.
\label{compact}
\ee
We take the $S^1$ to be large compared to the $T^4$, which we consider to be string-scale. We wrap $N_1$ D1 branes on $S^1$, and $N_5$ D5 branes on $S^1\times T^4$.

At low energies, the gauge theory on the bound state flows to a $(4,4)$ SCFT. It is conjectured that there is a point in moduli space where this SCFT is a symmetric product orbifold theory, consisting of $N_1N_5$ symmetrized copies of a free $c=6$ SCFT \cite{\orbifoldrefs}. 

We work in a Euclideanized theory where the base space is a cylinder, which we parameterize via: 
\be
w = \tau + i\sigma, \quad 0\le \sigma<2\pi, \quad-\infty<\tau<\infty \,.
\ee
The target space is the symmetrized product of $N_1N_5$ copies of $T^4$:
\be
(T_4)^{N_1N_5}/S_{N_1N_5} \,.
\ee
Each copy of $T^4$ gives 4 bosonic fields $X^1, X^2, X^3, X^4$, along with 4 left-moving fermionic excitations $\psi^1, \psi^2, \psi^3, \psi^4$ and the corresponding right-moving excitations, which we denote with a bar ($\bar{\psi}^1$, etc.).  The central charge of the theory on each copy of $T^4$ is $c=6$, which yields a total central charge of $6 N_1N_5$.

\subsection{Symmetries of the CFT}
The D1D5 CFT has a (small) $\mathcal{N}=4$ superconformal symmetry in both the left and right-moving sectors. The generators and their OPEs are given in Appendix~\ref{ap:CFT-notation}. 

Each superconformal algebra contains an R symmetry $SU(2)$.  Thus we have the global symmetry $SU(2)_L\times SU(2)_R$, with quantum numbers:
\be\label{ExternalQuantumNumbers}
SU(2)_L: ~(j, m);~~~~~~~SU(2)_R: ~ (\bar j, \bar m).
\ee
In addition there is an $SO(4) \simeq SU(2)\times SU(2)$ symmetry, coming from rotations in the four directions of the $T^4$, which is broken by the fact that we have compactified these directions into a torus.  However, the symmetry still provides a useful organizing tool for states. We label this symmetry by
\be
SU(2)_1\times SU(2)_2.
\ee

We use indices $\alpha, \dot\alpha$ for $SU(2)_L$ and $SU(2)_R$ respectively, and indices $A, \dot A$ for $SU(2)_1$ and $SU(2)_2$ respectively.  The 4 real fermion fields of the left sector are grouped into complex fermions $\psi^{\alpha A}$. The right fermions are grouped into fermions $\bar{\psi}^{\dot\alpha A}$. The boson fields $X^i$ are a vector in $T^4$ and have no charge under $SU(2)_L$ or $SU(2)_R$, so are grouped as $X_{A \dot A}$. Different copies of the $c=6$ CFT are denoted with a copy label in brackets, e.g.~
\bea
X^{(1)} \,, ~ X^{(2)} \,,~ \cdots \,, ~X^{(N_1N_5)} \,.
\eea
Further details are given in Appendix~\ref{ap:CFT-notation}.

\subsection{NS and R vacua}
Consider a single copy of the $c=6$ CFT. The lowest energy state of the left-moving sector is the NS vacuum, 
\be
|0_{NS}\rangle \,, \qquad\quad h = 0, \quad m = 0 \,
\ee
where $h$ denotes the $L_0$ eigenvalue.
However, we study the CFT in the R sector. In particular we are interested in the R vacua denoted by\footnote{There are two other left-moving R ground states, which may be obtained by acting with fermion zero modes. Including right-movers then gives a total of 16 R-R ground states; see e.g.~\cite{Avery:2010qw}.}
\be
|0_{R}^{\pm}\rangle \,, 
\qquad\quad h=\frac14, \quad m=\pm\frac12 \,.
\ee
We can relate the NS and R sectors using spectral flow.  In particular, spectral flow by $\a = \pm 1$ in the left-moving sector produces the transformations
\bea
\alpha= 1 : && \ket{0_R^-} \rightarrow \ket{0_{NS}} \,, 
\quad \ket{0_{NS}}  \rightarrow \ket{0_R^+}  \cr
\alpha= -1 : && \ket{0_R^+} \rightarrow \ket{0_{NS}} \,, 
\quad \ket{0_{NS}}  \rightarrow \ket{0_R^-}  \label{eq:sfstates}
\eea

Similar relations hold for the right-moving sector, which we denote with bars.  In many places we will write expressions only for the left-moving sector, as the two sectors are on an equal footing in this paper.  However, in certain places we will write expressions involving the full vacuum of both sectors.  We therefore introduce the notation
\bea
|0_R^{--}\rangle &\equiv& |0_R^-\rangle |\bar{0}_R^-\rangle \,. \label{eq:0r--}
\eea
Following the notation of \cite{\acmtwo}, we denote the hermitian conjugate of $\ket{0_R^{--}}$ as 
\bea
\bra{0_{R,--}} \,.
\eea

\subsection{Twist operators and twisted R vacua}\label{TwistOperators}

The symmetric orbifold theory contains twist operators, as well as more general spin-twist operators in the twisted sector.  Let us recall their definition and, in doing so, we will define the initial and final states of the amplitude we compute.  We will mostly follow the content of \cite {\lmone, \lmtwo}, but with a different presentation which is more appropriate for our purposes\footnote{See also the discussion in \cite{Avery:2009xr}.}. 

The bare twist operator $\s_M$ is defined as follows.  Let us work on the cylinder, and insert $\s_M$ at a point $w_*$. As the fields circle this point, the fields transform as
\bea
X^{(1)} \rightarrow X^{(2)} \rightarrow \cdots \rightarrow X^{(M)} \rightarrow X^{(1)} ~~ \label{eq:bp1} \\
\psi^{(1)} \rightarrow \psi^{(2)} \rightarrow \cdots \rightarrow \psi^{(M)} \rightarrow \; -\psi^{(1)}. \label{eq:fp1}
\eea
Note the minus sign in the fermionic expression.  These boundary conditions are achieved as follows.  First, we map the problem to the plane with coordinate $z$ via 
\bea
z=e^w
\eea
and then to a local covering plane with coordinate $t$ via a map which has the local form
\bea
z-z_* \approx b_* \left(t-t_* \right)^M
\label{eq:locmap}
\eea
where $z_*$ and $t_*$ are the respective images of the insertion point $w_*$ in the $z$ plane and the $t$ plane.  In the $t$ plane, we insert the identity operator at the point $t_*$. The $ M $ bosonic fields in \eq{eq:bp1} map to one single-valued bosonic field $X(t)$ in the $t$ plane, and similarly for the fermions.

In the absence of other insertions, we end up with an empty $t$ plane, i.e. in the NS vacuum. Equivalently, one can think in terms of the `covering cylinder' with coordinate $u$, defined by 
\bea
t=e^u.
\eea
In the NS vacuum on the $u$ cylinder, the fermions are antiperiodic, which agrees with \eq{eq:fp1}.

In the present paper we are interested in an initial state which involves component strings of length $M,N$ in R vacuum states.
Thus, we next define the (spin-)twist operator $\s^{\pm}_M$ as follows.  Follow the procedure used to define the bare twist $\s_M$, but in the covering $t$ plane, insert a spin field $S^{\pm}$ at $t_*$ (if $b_* \neq 1 $ in \eq{eq:locmap}, we also need to include a normalization factor, which we will ignore here, but which will be important in Section~\ref{sec:CMN}). If we take $t_*=0$, we obtain the R vacuum $|0_R^{\pm} \rangle _t$ of the $t$ plane. We write 
\bea
\s_M^{\a} = S_M^{\a} \sigma_M \,,   \qquad \a = +,- \,.
\eea
Back on the original cylinder, with coordinate $w$, as the fields circle the operator $\s_M^{\pm}$, they transform as 
\bea
X^{(1)} \rightarrow X^{(2)} \rightarrow \cdots \rightarrow X^{(M)} \rightarrow X^{(1)}\\
\psi^{(1)} \rightarrow \psi^{(2)} \rightarrow \cdots \rightarrow \psi^{(M)} \rightarrow +\psi^{(1)}.
\eea
Note the plus sign in the fermionic expression. On the covering cylinder with coordinate $u$, the fermion field is periodic, and if the operator is inserted at past infinity, we obtain the R vacuum $|0_R^{\pm} \rangle_u$ of the covering cylinder. 
We write the corresponding state on the original cylinder (with coordinate $w$) as $|0_R^\pm\rangle_M$. 

Adding in the right-moving sector, we obtain the full spin-twist field
\bea
\s_M^{\alpha \dot \alpha} = S_M^{\alpha} \bar{S}_M^{\dot \alpha} \s_M
\eea
and the corresponding state $|0_R^{\alpha \dot \alpha}\rangle_M$. 

We can now write our full initial state as $|0_R^{--}\rangle^{(1)}_M |0_R^{--}\rangle^{(2)}_N$. For ease of notation however,  we shall immediately suppress the subscripts $M$, $N$ on the states, thus writing our full initial state as
\bea
|0_R^{--}\rangle^{(1)} |0_R^{--}\rangle^{(2)}.
\eea
We shall denote states on the final component string of length $M + N $ without a copy label, and we will express the states produced by the action of the deformation operator in terms of excitations above the R-R ground state
\bea
|0_R^{--}\rangle \,.
\eea
For the remainder of the paper, $|0_R^{--}\rangle$ and its hermitian conjugate 
$\bra{0_{R,--}}$ will denote states on the final component string of length $M + N $.

\subsection{The deformation operator}

In the previous subsection, we defined the operators $\s_M^{\a\dot\a}$.
When $M=2$, the operator $\s_2^{++}$ is a chiral primary in both left- and right-moving sectors, with quantum numbers
\bea
h=j=m=\frac{1}{2}, \qquad \bar h =\bar j = \bar m =\frac{1}{2}. 
\eea
The twist operator $\sigma^{\a\dot\a}_2$ is normalized to have a unit OPE with its conjugate $\sigma_{2,{\a\dot\a}}$,
\be
\sigma_{2,{\a\dot\a}}(w,\bar{w})\sigma_2^{\a\dot\a}(w',\bar{w}')\sim {1\over |w-w'|^2}
\ee
where the above indices $\a$, $\dot\a$ are \emph{not} summed over.

The deformation operator is given by (see e.g.~the discussion in \cite{\acmtwo})
\be
\hat O_{\dot A\dot B}(w_0,\bar{w}_0)=
\left[{1\over 2\pi i} \int\limits_{w_0} dw G^-_{\dot A} (w)\right]
\left[{1\over 2\pi i} \int\limits_{\bar w_0} d\bar w 
\bar G^-_{\dot B} (\bar w)\right]
\sigma_2^{++}(w_0,\bar{w}_0)
\ee
The $\dot A, \dot B$ indices can be contracted to rewrite the above four operators as a singlet and a triplet of $SU(2)_2$.

\begin{figure}[t]
\begin{center}
\includegraphics[scale=.38]{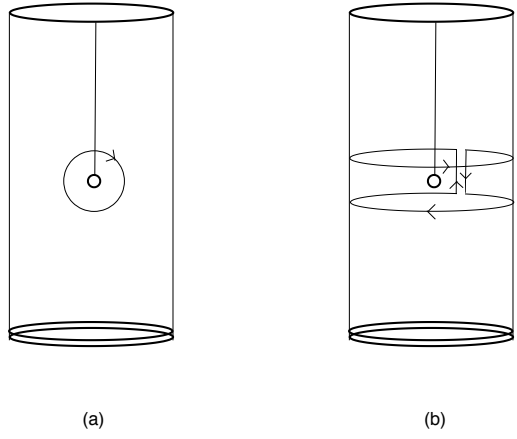}
\caption{The deformation operator involves the application of a supercharge, integrated around the twist operator.  The contour may be stretched as shown, so that the deformation operator may be written in terms of a combination of applying the supercharge before and after the twist.}
\label{ftwo}
\end{center}
\end{figure}

Let us now switch to writing only the holomorphic parts of states and operators, for ease of notation. 
The deformation operator involves the application of a supercharge $G^{-}$, which is integrated around the twist operator $\s_2^+$.  This can be written as a combination of applying the supercharge before and after the twist (see Fig.\;\ref{ftwo}); the upper part  of the contour gives
\begin{eqnarray}
\frac{1}{2\pi i}\int\limits_{w=\tau_0+\e}^{\tau_0+\e+2\pi i(M+N)}G_{\dot{A}}^{-}\left(w\right)dw &=& G_{\dot{A},0}^{-}
\label{G0_above}
\end{eqnarray}
and the lower 
 part  of the contour gives
\begin{eqnarray}
- \frac{1}{2\pi i} \left[\int\limits_{w=\tau_0-\e}^{\tau_0-\e+2\pi iM}G_{\dot{A}}^{(1)-}
\left(w\right)dw +\int\limits_{w=\tau_0-\e}^{\tau_0-\e+2\pi iN}G_{\dot{A}}^{(2)-}
\left(w\right)dw \right]
&=& - \left( G_{\dot{A},0}^{(1)-} +G_{\dot{A},0}^{(2)-} \right) . \qquad\quad
\label{G0_below-1}
\end{eqnarray}
Thus we obtain
\be
\hat{O}_{\dot A} = G^{-}_{\dot A,0}\s_2^+ - \s_2^+ \left ( G^{(1)-}_{\dot A,0} + G^{(2)-}_{\dot A,0}\right )
\ee
and similarly for the right sector.  When acting upon a state, we can use the fact that the supercharge annihilates the vacuum to simplify our expressions.  For the vacuum, we have
\bea
\hat{O}_{\dot A}|0_R^-\rangle^{(1)} |0_R^-\rangle^{(2)} & = & G^{-}_{\dot A,0}\s_2^+|0_R^-\rangle^{(1)} |0_R^-\rangle^{(2)}
\eea
while for some operator $\hat{Q}$, we have
\bea
\hat{O}_{\dot A}\hat{Q}|0_R^-\rangle^{(1)}  |0_R^-\rangle^{(2)} & = & G^{-}_{\dot A,0}\s_2^+\hat{Q}|0_R^-\rangle^{(1)}  |0_R^-\rangle^{(2)} - \s_2^+ \left [ \left ( G^{(1)-}_{\dot A,0} + G^{(2)-}_{\dot A,0}\right ),\hat{Q} \right ]|0_R^-\rangle^{(1)}  |0_R^-\rangle^{(2)} \nn
&=& G^{-}_{\dot A,0}\s_2^+\hat{Q}|0_R^-\rangle^{(1)}  |0_R^-\rangle^{(2)} - \s_2^+ \left \{ \left ( G^{(1)-}_{\dot A,0} + G^{(2)-}_{\dot A,0}\right ),\hat{Q} \right \}|0_R^-\rangle^{(1)}  |0_R^-\rangle^{(2)} \nn
\eea
The commutation/anticommutation relations of the supercharge with basic operators such as the creation and annihilation operators for the bosonic and fermionic fields are known.  As such, the bulk of our computations focus on the effects of the twist.

\section{Introducing the computation} \label{sec:intrcomp}

\subsection{Mode Expansions on the cylinder} \label{sec:expansions}

In order to set up the computation and present the explicit form of the exponential ansatz,  we now give the mode expansions of the fundamental fields on the cylinder. 
There are two component strings before the twist;  string (1) is wound $M$ times, while string (2) is wound $N$ times. 

Before the twist, the mode expansions are
\begin{eqnarray}
\a_{A\dot A ,m}^{(1)} &=& \frac{1}{2\pi i} \int\limits_{\sigma=0}^{2\pi M} \partial X^{(1)}_{A\dot A} (w) e^{\frac{m}{M}w} dw
\label{alphaM} \\
\a_{A\dot A , m}^{(2)} &=&  \frac{1}{2\pi i} \int\limits_{\sigma=0}^{2\pi N} \partial X^{(2)}_{A\dot A} (w) e^{\frac{m}{N}w} dw
\label{alphaN} \\
d_m^{(1)\alpha A} &=& \frac{1}{2\pi i\sqrt M} \int\limits_{\sigma=0}^{2\pi M} \psi^{(1)\alpha A} (w) e^{\frac{m}{M}w} dw
\label{fermM} \\
d_m^{(2)\alpha A} &=&  \frac{1}{2\pi i\sqrt N} \int\limits_{\sigma=0}^{2\pi N} \psi^{(2)\alpha A} (w) e^{\frac{m}{N}w} dw \,.
\label{fermN}
\eea
From the two-point functions, the commutation/anticommutation relations are
\bea
\left [ \a_{A\dot A , m}^{(i)}, \a_{B\dot B , n}^{(j)} \right ] &=& -m\e_{AB}\e_{\dot A \dot B}\delta^{ij}\delta_{m+n,0} \cr
\left\{d_m^{(i)\alpha A}, d_n^{(j)\beta B}\right\} &= &-\varepsilon ^{\alpha\beta} \varepsilon^{AB}\delta^{ij}\delta_{m+n,0} \,.
\label{commrelation}
\eea
After the twist, there is a single component string of length $ M + N $. We thus have modes
\bea
\a_{A\dot A ,m} &=& \frac{1}{2\pi i} \!\! \int\limits_{\sigma=0}^{2\pi (M+N)} \partial X_{A\dot A} (w) e^{\frac{m}{M+N}w} dw 
\label{alphaMpost} \\
d_k^{\alpha A} &=& \frac{1}{2\pi i\sqrt{M+N}} \!\! \int\limits_{\sigma=0}^{2\pi (M+N)} \psi^{\alpha A} (w) e^{\frac{k}{M+N}w} dw
\label{fermMN}
\eea
with commutation relations
\bea
\left [ \a_{A\dot A , m}, \a_{B\dot B , n} \right ] &=& -m\e_{AB}\e_{\dot A \dot B}\delta_{m+n,0} 
\label{eq:bcrat}
\\
\left\{d_k^{\alpha A}, d_l^{\beta B}\right\} &=& -\varepsilon^{\alpha\beta} \varepsilon^{AB} \delta_{k+l,0}  \,.
\eea

\subsection{The exponential ansatz}

We begin our computations by working firstly with the twist operator $\s^{++}_2$.  After determining the effects of the twist operator, we use the known commutation relations of the supercharge to determine the effects of the full deformation operator.  We consider separately the effects of this operator on the vacuum, and on a state with a single bosonic or fermionic excitation.

Let us now record the exponential ansatz with all notation explicit:
\begin{eqnarray}
\ket{\boldsymbol{\chi}} &\equiv& \sigma_{2}^{++}(w_{0})
|0_R^{--}\rangle^{(1)}|0_R^{--}\rangle^{(2)} 
\cr
&=&
C_{MN} \left[
\exp\left(\sum_{m\geq 1,n\geq 1}
\gamma_{mn}^{B}\left[-\alpha_{++,-m}\alpha_{--,-n}
+\alpha_{+-,-m}\alpha_{-+,-n}\right]\right)
\right. \cr
&& \qquad \qquad  \left.
\times\exp\left(\sum_{m\geq 0,n\geq 1}
\gamma_{mn}^{F}
\left[d_{-m}^{++}d_{-n}^{--}-d_{-m}^{+-}d_{-n}^{-+}
\right]\right)
\times [\mathrm{antihol.}] \right]
\ket{0_{R}^{--}} \,. \qquad
\label{eq:expans}
\end{eqnarray}
This ansatz is a generalization of that made in \cite{\acmtwo} in the case of $M=N=1$, where the exponential ansatz was verified up to fourth order in the bosonic oscillators.

The main novel feature of the above ansatz relative to the $M=N=1$ case is the non-trivial overall factor $C_{MN}$. A priori, this could be a function of $w_0$, but will turn out to be a pure c-number independent of $w_0$. We observe that
\bea
C_{MN}&=&\bra{0_{R,--}}
\sigma_{2}^{++}\left(w_{0}\right)
\ket{0_{R}^{--}}^{\left(1\right)}\ket{0_{R}^{--}}^{\left(2\right)} \,.
\label{eq:cmn}
\eea
In the case of $M=N=1$, the normalization of the twist $\sigma_2^{++}$ implies that $C_{11}=1$ \cite{\acmtwo}. 
We compute the general factor $C_{MN}$ in Section \ref{sec:CMN}.

In the case of an initial bosonic or fermionic excitation, the above exponential ansatz implies \cite{\acmthree,\cmt} that the twist operator converts an initial excitation into a linear combination of excitations above the state $\ket{\boldsymbol{\chi}}$, which we write as 
\bea
\sigma_{2}^{++}(w_{0})\alpha^{(i)}_{A\dot A,-m} |0^{--}_R\rangle^{(1)} |0^{--}_R\rangle^{(2)} ~&=&~\sum\limits_k f^{B(i)}_{mk}\,  \alpha_{A\dot A,-k}|\boldsymbol{\chi}\rangle \,, \qquad i=1,2
\label{eq:fbdef} \\
\sigma_{2}^{++}(w_{0})d^{(i)\pm A}_{-m} |0^{--}_R\rangle^{(1)} |0^{--}_R\rangle^{(2)} ~&=&~\sum\limits_k f^{F(i)\pm}_{mk}\,  d^{(i)\pm A}_{-k}|\boldsymbol{\chi}\rangle \,, \qquad i=1,2 \,.
\label{eq:ffdef}
\eea

\section{The method of computation} \label{computation}

\subsection{Mapping to a covering plane}

In order to relate states before the twist to states after the twist, we first map the cylinder to the complex plane via:
\be
z = e^w = e^{\tau+i\s} \,.
\ee
The point $w_0$ maps to a point $z_0$.
We next map the problem to a covering plane, with coordinate $ t $, via the map
\bea\label{eq:covermap}
z &=& t^M(t-a)^N \,.
\eea
This map contains an $M$ fold bifurcation at $t = 0$, an $N$ fold bifurcation at $t = a$, and an $M+N$ fold bifurcation at $t = \infty$ (see Fig.\;\ref{fthree}).  
The bosonic fields $X^{(1)}$, $X^{(2)}$ map to one single-valued bosonic field $X(t)$ in the $t$ plane, and similarly for the fermions. 
The pre-twist component string (1) with winding $M$ maps to contours around the origin in the $t$-plane, while the pre-twist component string (2) with winding $N$ maps to contours around the point $t = a$.  The post-twist component string of winding $M+N$ maps to contours around $t = \infty$.  

The map also contains a two-fold bifurcation point at the other solution of $\tfrac{dz}{dt} = 0$, which is the location of the twist $\s_2^{++}$.  It is convenient to write the derivative ${dz \over dt}$ as
\be
{dz \over dt} = (M+N){z \over t(t-a)}\left ( t - {M\over M+N}a \right ) \,.
\ee
We see that
\be
t_0 = {M\over M+N}a \,
\ee
which corresponds via \eq{eq:covermap} to the following value of $z$,
\be
z_0={a^{M+N}} {M^MN^N\over (M+N)^{M+N}}(-1)^N \,.
\ee
To invert this for $a$, we must specify how to deal with the fractional exponent. We choose
\be
z_0={a^{M+N}} {M^MN^N\over (M+N)^{M+N}}e^{i\pi N} \,
\ee
which determines $a$ in terms of $z_0=e^{w_0}$ to be
\be
a=e^{-i\pi \frac{N}{M+N}}\left( {z_0\over M^MN^N}\right)^{1\over M+N}(M+N)\,.
\label{az}
\ee

We will perform our calculations using the parameter $a$, and then use this relation to re-express our results in terms of the insertion point $z_0$.

\begin{figure}[t]
\begin{center}
\includegraphics[scale=.45]{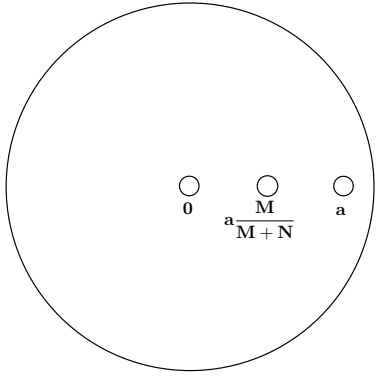}
\caption{The covering plane with coordinate $t$. The bifurcation points of the map are shown.}
\label{fthree}
\end{center}
\end{figure}

\subsection{Spectral flowing to an empty covering plane} \label{sec:sf}

In the $t$ plane, there are no twist insertions.  However, we have spin field insertions at each of the bifurcation points:
\begin{eqnarray}
S^- (t=0) \,,\quad S^+ (t = \tfrac{M}{M+N}a)\,,\quad S^- (t=a) \,, \quad S^+ (t=\infty)\,
\end{eqnarray}
along with appropriate normalization factors. 
To compute $\gamma^F$ and $f^{F(i)\pm}$, we shall take a ratio of amplitudes, and so it it will not be necessary to keep account of the normalization factors of the spin fields at this point; such factors will cancel out. 
However, these normalization factors will later be important when we come to calculate  the overall coefficient $C_{MN}$.

Our method of computing $\gamma^F$ and $f^{F(i)\pm}$ is to perform a sequence of spectral flows and coordinate transformations to
remove the spin fields, and thus to map the problem to an empty $ t $ plane.

The sequence is as follows (we write only the left-moving part, the right-moving part is identical):
\begin{enumerate}
\item[i)] Spectral flow by $\alpha = 1$ in the $t$ plane.

\item[ii)] Change coordinate to $t' = t - \tfrac{M}{M+N} a$.

\item[iii)] Spectral flow by $\alpha = -1$ in the $t'$ plane.

\item[iv)]  Change coordinate to $\hat t = t' - \frac{N}{M+N}a = t-a$.

\item[v)]  Spectral flow by $\alpha = 1$ in the $\hat t$ plane.

\item[vi)] Change coordinate back to $t = \hat t + a$ (when necessary).
\end{enumerate}

From the action of spectral flow on the R ground states in \eq{eq:sfstates} one sees that this sequence indeed maps the problem to one with no spin field insertions. 

The action of spectral flow on operators is straightforward for those operators where the fermion content may be expressed as a simple exponential in the language in which the fermions are bosonized. For such operators with charge $j$, spectral flow with parameter $\alpha$ gives rise to the transformation
$$\hat O_j (t)\to t^{-\alpha j}\hat O_j (t).$$
The fermion fields and spin fields are of this form.

Since we take a ratio of amplitudes to compute $\gamma^F$ and $f^{F(i)\pm}$, we will not need to take account of the transformation of the spin fields (again, these will however be important for the computation of $C_{MN}$).

By contrast, an essential part of the computation of $\gamma^F$ and $f^{F(i)\pm}$ is to follow the transformation of the modes of the fermion fields through the above sequence of spectral flows and coordinate transformations. 
Here we present only the final expressions for these modes; details of the derivation are given in Appendix \ref{sec:mapping}.

In terms of the $\hat t$-plane, the cylinder modes \eq{fermM}, \eq{fermN} and \eq{fermMN} map to the following expressions.  For the modes before the twist, we have
\begin{eqnarray}
    d_{m}^{\left(1\right)+A}&\to& \hat{d}_{m}'^{\left(1\right)+A}
		~=~\frac{\sqrt{M+N}}{2\pi i\sqrt{M}}
		\oint\limits_{\hat{t}=-a}d\hat{t}\,\psi^{\left(1\right)+A}\left(\hat{t}\;\!\right)\left[\left(\hat{t}+a\right)^{m-1}\left(\hat{t}+\tfrac{Na}{M+N}\right)\hat{t}^{\frac{Nm}{M}-1}\right]\cr
     d_{m}^{\left(2\right)+A}&\to& \hat{d}_{m}'^{\left(2\right)+A}
		~=~\frac{\sqrt{M+N}}{2\pi i\sqrt{N}}
		\oint\limits_{\hat{t}=0}d\hat{t}\,\psi^{\left(2\right)+A}\left(\hat{t}\;\!\right)\left[\left(\hat{t}+a\right)^{\frac{Mm}{N}-1}\left(\hat{t}+\tfrac{Na}{M+N}\right)\hat{t}^{m-1}\right]\cr
     d_{m}^{\left(1\right)-A}&\to& \hat{d}_{m}'^{\left(1\right)-A}
		~=~\frac{\sqrt{M+N}}{2\pi i\sqrt{M}}
		\oint\limits_{\hat{t}=-a}d\hat{t}\,\psi^{\left(1\right)-A}\left(\hat{t}\;\!\right)\left[\left(\hat{t}+a\right)^{m}\hat{t}^{\frac{Nm}{M}}\right]\cr
     d_{m}^{\left(2\right)-A}&\to& \hat{d}_{m}'^{\left(2\right)-A}
		~=~\frac{\sqrt{M+N}}{2\pi i\sqrt{N}}
		\oint\limits_{\hat{t}=0}d\hat{t}\,\psi^{\left(2\right)-A}\left(\hat{t}\;\!\right)\left[\left(\hat{t}+a\right)^{\frac{Mm}{N}}\hat{t}^{m}\right]
\end{eqnarray}
After the twist, we obtain
\begin{eqnarray}
   d_{k}^{+A}&\to&\hat{d}_{k}'^{+A}~=~\frac{1}{2\pi i}\oint\limits_{\hat{t}=\infty}d\hat{t}\,\psi^{+A}\left(\hat{t}\;\!\right)\left[\left(\hat{t}+a\right)^{\tfrac{Mk}{M+N}-1}\left(\hat{t}+\tfrac{N}{M+N}a\right)\hat{t}^{\frac{Nk}{M+N}-1}\right]\cr
   d_{k}^{-A}&\to&\hat{d}_{k}'^{-A}~=~\frac{1}{2\pi i}\oint\limits_{\hat{t}=\infty}d\hat{t}\,\psi^{-A}\left(\hat{t}\;\!\right)\left[\left(\hat{t}+a\right)^{\tfrac{Mk}{M+N}}\hat{t}^{\frac{Nk}{M+N}}\right]
\label{dhat}
\end{eqnarray}

For later use, let us also rewrite these modes in the $t$ plane.  The modes before the twist then become
\begin{eqnarray}
    d_{m}^{\left(1\right)+A}&\to& d_{m}'^{\left(1\right)+A}~=~\frac{\sqrt{M+N}}{2\pi i\sqrt{M}}\oint\limits_{t=0}dt\,\psi^{\left(1\right)+A}\left(t\right)\left[t^{m-1}\left(t-\tfrac{M}{M+N}a\right)\left(t-a\right)^{\frac{Nm}{M}-1}\right]\cr
     d_{m}^{\left(2\right)+A}&\to& d_{m}'^{\left(2\right)+A}~=~\frac{\sqrt{M+N}}{2\pi i\sqrt{N}}\oint\limits_{t=a}dt\,\psi^{\left(2\right)+A}\left(t\right)\left[t^{\frac{Mm}{N}-1}\left(t-\tfrac{M}{M+N}a\right)\left(t-a\right)^{m-1}\right]\cr
     d_{m}^{\left(1\right)-A}&\to& d_{m}'^{\left(1\right)-A}~=~\frac{\sqrt{M+N}}{2\pi i\sqrt{M}}\oint\limits_{t=0}dt\,\psi^{\left(1\right)-A}(t)\left[t^{m}\left(t-a\right)^{\frac{Nm}{M}}\right]\cr
      d_{m}^{\left(2\right)-A}&\to& d_{m}'^{\left(2\right)-A}~=~\frac{\sqrt{M+N}}{2\pi i\sqrt{N}}\oint\limits_{t=a}dt\,\psi^{\left(2\right)-A}(t)\left[t^{\frac{Mm}{N}}\left(t-a\right)^{m}\right]
\label{d_t(i)_modes}
\end{eqnarray}
and after the twist we obtain
\begin{eqnarray}
   d_{k}^{+A}&\to&d_{k}'^{+A}~=~\frac{1}{2\pi i}\oint\limits_{t=\infty}dt\,\psi^{+A}\left(t\right)\left[t^{\tfrac{Mk}{M+N}-1}\left(t-\tfrac{M}{M+N}a\right)\left(t-a\right)^{\frac{Nk}{M+N}-1}\right]\cr
   d_{k}^{-A}&\to&d_{k}'^{-A}~=~\frac{1}{2\pi i}\oint\limits_{t=\infty}dt\,\psi^{-A}\left(t\right)\left[t^{\tfrac{Mk}{M+N}}\left(t-a\right)^{\frac{Nk}{M+N}}\right] .
 \label{d_t_modes}
\end{eqnarray}
For later use, we also define modes natural to the $t$ and $\hat{t}$ planes,
\begin{eqnarray}
\tilde{d}_{r}^{\alpha A}&=&\frac{1}{2\pi i}\oint\limits_{t=0}\psi^{\alpha A}(t)t^{r-\frac{1}{2}} \label{nat_t_modes} \\
\hat{d}_{r}^{\alpha A}
&=&\frac{1}{2\pi i}\oint\limits_{\hat t=0}
\psi^{\alpha A}\left(\hat{t}\;\!\right){\hat t}^{r-\frac{1}{2}} \,.
\label{nat_that_modes}
\end{eqnarray}
The anticommutation relations for these modes are
\begin{eqnarray}
\lbrace \tilde{d}^{\alpha A}_{r},\tilde{d}^{\beta B}_{s}\rbrace~=~\lbrace \hat{d}^{\alpha A}_{r},\hat{d}^{\beta B}_{s}\rbrace&=&-\epsilon^{\alpha\beta}\epsilon^{AB}\delta_{r+s,0} \,.
\label{anti_comm}
\end{eqnarray}

\section{Effect of the deformation operator on the vacuum} \label{sec:gamma}

We now find the effect of the deformation operator on the R-R ground state 
$\ket{0_{R}^{--}}^{\left(1\right)} \ket{0_{R}^{--}}^{\left(2\right)}$.
We start by computing the effect of the twist operator $\s_2^{++}$, and we then apply the supercharge. 

The effect of the twist operator $\s_2^{++}$ on the state 
$\ket{0_{R}^{--}}^{\left(1\right)} \ket{0_{R}^{--}}^{\left(2\right)}$ is described in terms of the quantities $\g^B$, $\g^F$ and $C_{MN}$. The quantity $\g^B$ was computed in \cite{\chmtone} (and we will recall it later); we now proceed to calculating $\g^F$ and $C_{MN}$.

\subsection{Computing the Bogoliubov coefficients $\gamma^{F}$}

We now compute $\gamma_{kl}^{F}$. Let us consider the ratio of amplitudes
\begin{eqnarray}
\frac{\mathcal{A}_{2}}{\mathcal{A}_{1}}&=&\frac{\bra{0_{R,--}}\left(d^{++}_{l}d^{--}_{k}\right)\sigma_{2}^{+}\left(w_{0}\right)
\ket{0_{R}^{--}}^{\left(1\right)}
\ket{0_{R}^{--}}^{\left(2\right)}}
{\bra{0_{R,--}}\sigma_{2}^{+}\left(w_{0}\right)\ket{0_{R}^{--}}^{\left(1\right)}\ket{0_{R}^{--}}^{\left(2\right)}} \,.
\label{amp}
\end{eqnarray}
From the exponential ansatz \eq{eq:expans} we observe that
\begin{eqnarray}
\mathcal{A}_{2}&=&\bra{0_{R,--}}d^{++}_{l}d^{--}_{k}\ket{\boldsymbol{\chi}} \cr
&=& C_{MN} \bra{0_{R,--}}d^{++}_{l}d^{--}_{k} 
\exp\left(\sum_{m\geq 0,n\geq 1}
\gamma_{mn}^{F}
\left[d_{-m}^{++}d_{-n}^{--}-d_{-m}^{+-}d_{-n}^{-+}
\right]\right)
\ket{0_{R}^{--}} \cr
&=& C_{MN} \, \gamma_{kl}^{F}
\end{eqnarray}
and so using \eq{eq:cmn} we see that
\begin{eqnarray}
\gamma_{kl}^{F}&=&\frac{\mathcal{A}_{2}}{\mathcal{A}_{1}} \,.
\label{amp_ratio}
\end{eqnarray}
We now map this ratio of amplitudes to the empty $\hat t$ plane using the sequence of spectral flows and coordinate transformations discussed in Section \ref{sec:sf}, obtaining
\begin{eqnarray}
\gamma_{kl}^{F} ~=~ \frac{\bra{0_{R,--}}\left(d^{++}_{l}d^{--}_{k}\right)\sigma_{2}^{+}\left(w_{0}\right)\ket{0_{R}^{--}}^{\left(1\right)}\ket{0_{R}^{--}}^{\left(2\right)}}
{\bra{0_{R,--}}\sigma_{2}^{+}\left(w_{0}\right)\ket{0_{R}^{--}}^{\left(1\right)}\ket{0_{R}^{--}}^{\left(2\right)}}
&=&\frac{{}_{\hat{t}}\bra{0_{NS}}\hat{d}'^{++}_{l}\hat{d}'^{--}_{k}\ket{0_{NS}}_{\hat{t}}}
{{}_{\hat{t}}\langle {} 0_{NS}|0_{NS}\rangle_{\hat{t}}}
\label{amp-2}
\end{eqnarray}
where $\hat{d}'^{++}_l$ and $\hat{d}'^{--}_k$ are the modes after all spectral flows and coordinate transformations given in \eq{dhat}. 

Let us now evaluate the $\hat{t}$ plane amplitude in the numerator of (\ref{amp-2}). 
To do so, we expand the transformed modes $\hat{d}'$ in terms of the natural modes $\hat{d}$ on the $\hat{t}$ plane \eq{nat_that_modes}, whose commutation relations are known. From \eq{dhat} we have 
\begin{eqnarray}
\hat{d}_{l}'^{++}~=~\frac{1}{2\pi i}\oint\limits_{\hat{t}=\infty}d\hat{t}\,\psi^{++}\left(\hat{t}\;\!\right)\left[\left(\hat{t}+a\right)^{\tfrac{Ml}{M+N}-1}\left(\hat{t}+\tfrac{N}{M+N}a\right)\hat{t}^{\frac{Nl}{M+N}-1}\right]
\end{eqnarray}
Since we are at large $\hat{t}$ we expand
\begin{eqnarray}
\left(\hat{t}+a\right)^{\frac{Ml}{M+N}-1}\hat{t}^{\frac{Nl}{M+N}-1}
&=&
\hat{t}^{l-2}\left(1+a\hat{t}^{-1}\right)^{\frac{Ml}{M+N}-1}
~=~
\sum_{q\geq 0}{}^{\frac{Ml}{M+N}-1}C_{q} \, a^{q} \, \hat{t}^{l-q-2}
\label{binom_exp}
\end{eqnarray}
where ${}^x C_y$ is the binomial coefficient.  This gives
\begin{eqnarray}
\hat{d}_{l}'^{++}~=~
\sum_{q\geq 0}{}^{\frac{Ml}{M+N}-1}C_{q} 
\left( a^{q} \, \hat{d}_{l-q-\frac{1}{2}}^{++} 
+ \tfrac{N}{M+N} \, a^{q+1} \, \hat{d}_{l-q-\frac{3}{2}}^{++} \right) \,.
\end{eqnarray}
Similarly, for $\hat{d}'^{--}_k$ we obtain the relation
\begin{eqnarray}
\hat{d}_{k}'^{--}~=~
\sum_{p\geq 0}{}^{\frac{Mk}{M+N}}C_{p} \,
 a^{p} \, \hat{d}_{k-p+\frac{1}{2}}^{--} \,.
\end{eqnarray}
We then obtain 
\begin{eqnarray}
&&{}_{\hat{t}}\bra{0_{NS}}d'^{++}_{l}d'^{--}_{k}\ket{0_{NS}}_{\hat{t}}\cr
\cr
&&{}\qquad=~\sum_{p\geq 0}\sum_{q\geq 0}a^{p+q}\,{}^{\frac{Mk}{M+N}}C_{p}{}^{\frac{Ml}{M+N}-1}C_{q}\,\,{}_{\hat{t}}\bra{0_{NS}}\hat{d}^{++}_{l-q-\frac{1}{2}}\hat{d}^{--}_{k-p+\frac{1}{2}}\ket{0_{NS}}_{\hat{t}}\cr
&&{}\qquad\qquad+\frac{N}{M+N}\sum_{p\geq 0}\sum_{q\geq 0}a^{p+q+1}\,{}^{\frac{Mk}{M+N}}C_{p}{}^{\frac{Ml}{M+N}-1}C_{q}\,\,{}_{\hat{t}}\bra{0_{NS}}\hat{d}^{++}_{l-q-\frac{3}{2}}\hat{d}^{--}_{k-p+\frac{1}{2}}\ket{0_{NS}}_{\hat{t}} \qquad\quad
\label{t_hat_amp_binom}
\end{eqnarray}
We have two terms. For the first term, utilizing commutation relations \eq{anti_comm} we obtain
\begin{eqnarray}
l-q-\frac12=-\left(k-p+\frac12\right)&\Rightarrow&p=k+l-q \,.
\end{eqnarray}
Similarly, for the second term we obtain
\begin{eqnarray}
p=k+l-q-1 \,.
\end{eqnarray}
Note that in the first term, $\hat{d}^{++}_{l-q-\frac{1}{2}}$ must be an annihilation operator, so we have nonzero contributions only from modes with
\begin{eqnarray}
q \leq l-1
\end{eqnarray}
and similarly for the second term we have nonzero contributions only from
\begin{eqnarray}
q \leq l-2 \,.
\end{eqnarray}
Then (\ref{t_hat_amp_binom}) becomes
\begin{eqnarray}
&&{\phantom{\rangle}}_{\hat{t}}\bra{0_{NS}}d'^{++}_{l}d'^{--}_{k}\ket{0_{NS}}_{\hat{t}}
~=~ \cr
&& \quad
-a^{k+l} \left[
\sum_{q=0}^{l-1}{}^{\frac{Ml}{M+N}-1}C_{q}{}^{\frac{Mk}{M+N}}C_{l+k-q} 
+\frac{N}{M+N}\sum_{q=0}^{l-2}{}^{\frac{Mk}{M+N}-1}C_{q}{}^{\frac{Ml}{M+N}}C_{l+k-q-1}
\right]
{}_{\hat{t}}\langle {} 0_{NS}|0_{NS}\rangle_{\hat{t}} \qquad\qquad\qquad {} \nnm 
\end{eqnarray}
which gives
\begin{eqnarray}
\gamma_{kl}^{F}&=&
-a^{k+l} \left[
\sum_{q=0}^{l-1}{}^{\frac{Ml}{M+N}-1}C_{q}{}^{\frac{Mk}{M+N}}C_{l+k-q} 
+\frac{N}{M+N}\sum_{q=0}^{l-2}{}^{\frac{Mk}{M+N}-1}C_{q}{}^{\frac{Ml}{M+N}}C_{l+k-q-1}
\right]. \qquad 
\end{eqnarray}
Evaluating the sums in $Mathematica$ we find 
\begin{eqnarray}
\gamma_{kl}^{F}&=&\frac{a^{k+l}
}
{\pi^{2}}\sin\left[\tfrac{N\pi k}{M+N}\right]
\sin\left[\tfrac{N\pi l}{M+N}\right]
\frac{MN}{(M+N)^{2}}\frac{k}{k+l}\frac{\Gamma\left[\frac{Mk}{M+N}\right]\Gamma\left[\frac{Nk}{M+N}\right]}{\Gamma\left[k\right]}\frac{\Gamma\left[\frac{Ml}{M+N}\right]\Gamma\left[\frac{Nl}{M+N}\right]}{\Gamma\left[l\right]} \,. \cr &&
\end{eqnarray}

\subsubsection{Writing $\gamma^{F}$ in final form}

We next write $\gamma^{F}$ in final form by making the following changes. 
\begin{enumerate}
\item[(i)] We replace the parameter $a$ by $z_{0}=e^{w_0}$, so that our result is expressed in terms of the insertion point of the twist, using \eq{az}.
\item[(ii)] We define fractional modes:
\begin{eqnarray}
s=\frac{k}{M+N},&&s'=\frac{l}{M+N} \,.
\end{eqnarray}
The parameters $s$ and $s'$ then give directly the physical wavenumbers of the modes on the cylinder with coordinate $w$.
\item[(iii)]We define the shorthand notation
\begin{eqnarray}
\mu_{s} &\equiv& 1-e^{2\pi i Ms} \,.  
\end{eqnarray}
 Note that
 \begin{eqnarray}
 \sin\left[\frac{\pi N k}{M+N}\right]
=\sin\left(\pi Ns\right)
=-\frac{i}{2}e^{i\pi Ns}\left(1-e^{-2\pi i Ns}\right)
=-\frac{i}{2}e^{i\pi Ns}\mu_{s} \,.
 \end{eqnarray}
 \end{enumerate}
In this notation, $\gamma^F$ becomes 
\begin{eqnarray}
\tilde{\gamma}_{ss'}^{F}=-\frac{1}{4\pi^{2}}z_{0}^{s+s'}\mu_{s}\mu_{s'}\frac{s}{s+s'}\frac{MN}{(M+N)^{2}}\left(\frac{(M+N)^{M+N}}{M^{M}N^{N}}\right)^{s+s'}\frac{\Gamma\left[Ms\right]\Gamma\left[Ns\right]}{\Gamma\left[(M+N)s\right]}\frac{\Gamma\left[Ms'\right]\Gamma\left[Ns'\right]}{\Gamma\left[(M+N)s'\right]}. \cr
\label{gen_gamma}
\end{eqnarray}
In order to verify that in the case of $M=N=1$ this expression agrees with that computed in \cite{\acmtwo}, note that due to different choices in normalization of the modes of the fermions, the expression $\frac{1}{2}\gamma^{F}$ in this paper should agree with the $\gamma^{F}$ in \cite{\acmtwo}. One can check that this is indeed the case.

\subsection{Computing the overall prefactor $C_{MN}$} \label{sec:CMN}

We next compute the overall prefactor $C_{MN}$. Recall that from the exponential ansatz \eq{eq:expans} we have 
\bea
C_{MN}&=&\bra{0_{R,--}}
\sigma_{2}^{++}\left(w_{0}\right)
\ket{0_{R}^{--}}^{\left(1\right)}\ket{0_{R}^{--}}^{\left(2\right)} \,.
\label{eq:cmn-2}
\eea
We use the methods developed in \cite{\lmone,\lmtwo} to compute this correlator.
The full calculation is somewhat lengthy and is presented in Appendix \ref{app:cmn}; here we summarize the main steps.  
\begin{enumerate}
\item[(i)] We first lift to the $z$ plane. Since $\sigma_{2}^{++}$ has weight (1/2,1/2), we obtain the Jacobian factor contribution
\begin{eqnarray}
\left|\frac{dz}{dw}\right|_{z=z_{0}}&=&|a|^{M+N}\frac{M^{M}N^{N}}{(M+N)^{M+N}}
\end{eqnarray}
\item[(ii)] We compute the above correlator of spin-twist fields following the method of \cite{\lmone,\lmtwo} as follows. We work in a path integral formalism, and we define regularized spin-twist operators by cutting circular holes in the $ z $ plane.  
We lift to the covering space $ t $ where the fields become single-valued. 
In the $ t $ plane there is a non-trivial metric;  we take account of this by defining
a fiducial metric and computing the Liouville action. The Liouville action terms give
\begin{eqnarray}
2^{-\frac{5}{4}}
|a|^{-\frac{3}{4}\left(M+N\right)+\frac{1}{2}\left(\frac{M}{N}+\frac{N}{M}+1\right)}M^{-\frac{3}{4}M-\frac{1}{4}}N^{-\frac{3}{4}N-\frac{1}{4}}(M+N)^{\frac{3}{4}(M+N)-\frac{1}{4}} \,.
\end{eqnarray}
\item[(iii)] In the covering space, we must insert spin fields, each within an appropriate normalization to take account of the local form of the map near each insertion, $\left(z-z_{*}\right)\approx b_*\left(t-{t_{*}}\right)^{n}$ \cite{\lmtwo}. These normalization factors give the contribution
\begin{eqnarray}
&&|b_{\infty}|^{-\frac{1}{2(M+N)}}|b_{t_{0}}|^{-\frac{1}{4}}|b_{a}|^{-\frac{1}{2N}}|b_{0}|^{-\frac{1}{2M}}=\cr
&&\qquad\qquad
2^{\frac{1}{4}}|a|^{-\frac{1}{2}\left(\frac{M}{N}+\frac{N}{M}+\frac{M}{2}+\frac{N}{2}-1\right)}M^{-\frac{1}{4}(M-1)}N^{-\frac{1}{4}(N-1)}(M+N)^{\frac{1}{4}(M+N-3)} \,. \qquad\qquad 
\end{eqnarray}
\item[(iv)] Finally, the correlator of the spin fields in the $t$ plane gives
\bea
\frac{\langle S^{+}(\infty)S^{+}(t_{0})S^{-}(a)S^{-}(0)\rangle}{\langle S^{+}(\infty)S^{-}(0)\rangle}&=&\frac{(M+N)^2}{MN} \frac{1}{|a|}\,.
\label{eq:spinfieldans1}
\end{eqnarray}
\end{enumerate}
These four results combine to give the final result
\begin{eqnarray}
C_{MN}=\frac{M+N}{2MN} \,.
\end{eqnarray}

In order to write the full expression for the effect of the twist operator 
$\sigma_{2}^{++}$ on the state $\ket{0_{R}^{--}}^{\left(1\right)} \ket{0_{R}^{--}}^{\left(2\right)}$, let us recall the expression for $\gamma^{B}$ computed in \cite{\chmtone}, 
\bea
\gamma^B_{kl}=-
{(-a)^{k+l}\over \pi^2}
\sin\left[
\tfrac{\pi  Mk}{M+N} \right]
\sin\left[
\tfrac{\pi Ml}{M+N} \right]
{MN\over (M+N)^2}
{1\over k+l}
{\Gamma[{Mk\over M+N}]
\Gamma[{Nk\over M+N}]\over \Gamma[k]}
{\Gamma[{Ml\over M+N}]
\Gamma[{Nl\over M+N}]\over \Gamma[l]} \cr &&
\eea
which since $k$ and $l$ are integers can be rewritten using 
\begin{eqnarray}
(-1)^{k}\sin\left[\frac{\pi M k}{M+N}\right]&=&-\sin\left[\frac{\pi N k}{M+N}\right],
\label{sin}
\end{eqnarray}
which gives
\bea
\gamma^B_{kl}=-
{a^{k+l}\over \pi^2}
\sin\left[
\tfrac{\pi  Nk}{M+N} \right]
\sin\left[
\tfrac{\pi Nl}{M+N} \right]
{MN\over (M+N)^2}
 \, {1\over k+l} \,
{\Gamma[{Mk\over M+N}]
\Gamma[{Nk\over M+N}]\over \Gamma[k]}
{\Gamma[{Ml\over M+N}]
\Gamma[{Nl\over M+N}]\over \Gamma[l]} . \quad
\label{eq:gammaB}
\eea
Therefore, the full effect of the twist operator 
$\sigma_{2}^{++}(w_{0},\bar{w}_{0})$ 
on the state $\ket{0_{R}^{--}}^{\left(1\right)} \ket{0_{R}^{--}}^{\left(2\right)}$ is
\begin{eqnarray}
\ket{\boldsymbol{\chi}} &\equiv& \sigma_{2}^{++}(w_{0})
|0_R^{--}\rangle^{(1)}|0_R^{--}\rangle^{(2)} 
\cr
&=&
C_{MN} \left[
\exp\left(\sum_{m\geq 1,n\geq 1}
\gamma_{mn}^{B}\left[-\alpha_{++,-m}\alpha_{--,-n}
+\alpha_{+-,-m}\alpha_{-+,-n}\right]\right)
\right. \cr
&& \qquad \qquad  \left.
\times\exp\left(\sum_{m\geq 0,n\geq 1}
\gamma_{mn}^{F}
\left[d_{-m}^{++}d_{-n}^{--}-d_{-m}^{+-}d_{-n}^{-+}
\right]\right)
\times [\mathrm{antihol.}] \right]
\ket{0_{R}^{--}} \qquad
\label{eq:expans-2}
\end{eqnarray}
where
\bea
C_{MN}&=&\frac{M+N}{2MN} \cr
\gamma^B_{kl}&=&-
{a^{k+l}\over \pi^2}
\sin\left[
\tfrac{\pi  Nk}{M+N} \right]
\sin\left[
\tfrac{\pi Nl}{M+N} \right]
{MN\over (M+N)^2}
 \, {1\over k+l} \,
{\Gamma[{Mk\over M+N}]
\Gamma[{Nk\over M+N}]\over \Gamma[k]} \,
{\Gamma[{Ml\over M+N}]
\Gamma[{Nl\over M+N}]\over \Gamma[l]} \cr
\gamma_{kl}^{F}&=& \phantom{-}
\frac{a^{k+l}}{\pi^{2}}
\sin\left[
\tfrac{\pi  Nk}{M+N} \right]
\sin\left[
\tfrac{\pi Nl}{M+N} \right]
\frac{MN}{(M+N)^{2}}
 \, \frac{k}{k+l} \,
{\Gamma[{Mk\over M+N}]
\Gamma[{Nk\over M+N}]\over \Gamma[k]} \,
{\Gamma[{Ml\over M+N}]
\Gamma[{Nl\over M+N}]\over \Gamma[l]}  \,. \cr &&
\eea

\subsection{Applying the Supercharge}

We now apply the supercharge to obtain the full effect of the deformation operator on the state $|0_R^{--}\rangle^{(1)}|0_R^{--}\rangle^{(2)} $.
For ease of notation, we introduce notation for the holomorphic and antiholomorphic parts of the state $\ket{\boldsymbol{\chi}}$,
\bea
\ket{\boldsymbol{\chi}} &=& \ket{\chi}\ket{\bar\chi},
\eea
where we divide the prefactor $C_{MN}$ equally between the holomorphic and antiholomorphic parts, i.e.
\bea
\ket{\chi} &=& 
\sqrt{C_{MN}} \left[
\exp\left(\sum_{m\geq 1,n\geq 1}
\gamma_{mn}^{B}\left[-\alpha_{++,-m}\alpha_{--,-n}
+\alpha_{+-,-m}\alpha_{-+,-n}\right]\right)
\right. \cr
&& \qquad \qquad  \left.
\times\exp\left(\sum_{m\geq 0,n\geq 1}
\gamma_{mn}^{F}
\left[d_{-m}^{++}d_{-n}^{--}-d_{-m}^{+-}d_{-n}^{-+}
\right]\right)
\right]
\ket{0_{R}^{-}}. \qquad
\label{eq:expans-3}
\end{eqnarray}
Similarly, we write the final state obtained by acting with holomorphic and antiholomorphic supercharges as 
\bea
\ket{\Psi} &=& \ket{\psi}\ket{\bar\psi}
\eea
where
\begin{eqnarray} \label{psistart}
\ket{\psi}=G_{\dot{A},0}^{-}\ket{\chi} \,
\end{eqnarray}
and similarly for the antiholomorphic part.
We now compute the state $ \ket{\psi}$.

From \eq{G0_above} we have
\begin{eqnarray}
G_{\dot{A},0}^{-} &=& \frac{1}{2\pi i}\int\limits_{w=\tau_0+\e}^{w=\tau_0+\e+2\pi i(M+N)}G_{\dot{A}}^{-}\left(w\right)dw ~=~ \frac{i}{\sqrt{M+N}}\sum_{n=-\infty}^{\infty}d_{n}^{-A}\alpha_{A\dot{A},-n} \,.
\label{G0_above-1}
\end{eqnarray}
We wish to write (\ref{psistart}) with only negative index modes acting on $\ket{0_{R}^{-}}$. We thus write
\begin{eqnarray}
G_{\dot{A},0}^{-}&=&\frac{i}{\sqrt{M+N}}\left (\sum_{l>0}^{\infty}d^{-A}_{-l}\alpha_{A\dot{A},l}+\sum_{l>0}^{\infty}d_{l}^{-A}\alpha_{A\dot{A},-l} + d^{-A}_0 \a_{A \dot A, 0} \right )
\end{eqnarray}
We note e.g.~from \eq{eq:gammaB} that $\gamma^{B}_{kl}$ is symmetric, so we have 
\begin{eqnarray}
\frac{i}{\sqrt{M+N}}\sum_{l\geq 1}^{\infty}d^{-A}_{-l}\alpha_{A\dot{A},l}\ket{\chi}&=&\frac{i}{\sqrt{M+N}}\sum_{l\geq 1}\sum_{l\geq 1}l\gamma^{B}_{kl}d^{-A}_{-l}\alpha_{A\dot{A},-k}\ket{\chi}\cr
\frac{i}{\sqrt{M+N}}\sum_{k\geq 1}^{\infty}d_{k}^{-A}\alpha_{A\dot{A},-k}\ket{\chi}&=&-\frac{i}{\sqrt{M+N}}\sum_{k\geq 1}\sum_{l\geq 1}\gamma_{kl}^{F}d_{-l}^{-A}\alpha_{A\dot{A},-k}\ket{\chi} \cr
\frac{i}{\sqrt{M+N}}d^{-A}_{0}\alpha_{A\dot{A}0}\ket{\chi}&=& 0
\end{eqnarray}
where we have used the commutation relations in (\ref{commrelation}). Thus
\begin{eqnarray}
G_{\dot{A},0}^{-}\ket{\chi}&=&\frac{i}{\sqrt{M+N}}\sum_{k\geq 1,l\geq 1}\left(l\gamma^{B}_{kl}-\gamma_{kl}^{F}\right)d_{-l}^{-A}\alpha_{A\dot{A},-k}\ket{\chi} \,.
\end{eqnarray}
We observe that the $l$ and $k$ sums factorize and we obtain
\begin{eqnarray}
\ket{\psi} ~=~G_{\dot{A},0}^{-}\ket{\chi}&=&-\frac{i}{\pi^{2}}\frac{MN}{(M+N)^{\frac{5}{2}}}\left(\sum_{l\geq 1}a^{l}\sin\left[\tfrac{N\pi l}{M+N}\right]\frac{\Gamma\left[\frac{Ml}{M+N}\right]\Gamma\left[\frac{Nl}{M+N}\right]}{\Gamma\left[l\right]}d_{-l}^{-A}\right)\cr
&& \qquad\times\left(\sum_{k\geq 1}a^{k}\sin\left[\tfrac{N\pi k}{M+N}\right]\frac{\Gamma\left[\frac{Mk}{M+N}\right]\Gamma\left[\frac{Nk}{M+N}\right]}{\Gamma\left[k\right]}\alpha_{A\dot{A},-k}\right)\ket{\chi} \,. \qquad
\label{eq:psians}
\end{eqnarray}
One can check\footnote{In order to verify this, note that due to different choices in normalization of the modes of the fermions, one should replace $d^{-A}\to \frac{1}{\sqrt{2}}d^{-A}$; in addition, there is an overall minus sign due to the different directionality of the contour for $G^{-}_{\dot{A},0}$.} that in the case of $M=N=1$ this expression agrees with that computed in \cite{\acmtwo}.

Analogous expressions hold for 
\begin{eqnarray}
\ket{\bar{\psi}}&=&\bar{G}_{\dot{B},0}^{-}\ket{\bar{\chi}} \,,
\end{eqnarray}
and the complete final state is given by
\begin{eqnarray}
\ket{\Psi}=\ket{\psi} \ket{\bar{\psi}} \,.
\end{eqnarray}

\section{Effect of the deformation operator on excited states} \label{sec:f}

We now consider the case where one of the initial component strings has an initial oscillator excitation. We first consider the fermionic excitations.  From \eq{eq:ffdef} we recall the definition
\bea
\sigma_{2}^{++}(w_{0})d^{(i)\pm A}_{-m} |0^{--}_R\rangle^{(1)} |0^{--}_R\rangle^{(2)} ~&=&~\sum\limits_k f^{F(i)\pm}_{mk}\,  d^{(i)\pm A}_{-k}|\boldsymbol{\chi}\rangle \,, \qquad i=1,2 \,.
\label{eq:ffdef-2}
\eea
In this section we compute $f_{mk}^{F(i)\pm}$.

\subsection{Computing $f_{mk}^{F(1)+}$}

Let us start with $f_{mk}^{F(1)+}$. Consider the ratio of amplitudes
\begin{eqnarray}
\frac{\mathcal{A}_{3}}{\mathcal{A}_{1}}&=&\frac{
\bra{0_{R,--}}d_{k}^{--}\sigma_{2}^{++}\left(w_{0}\right)d_{-m}^{(1)++}\ket{0_{R}^{--}}^{(1)} \ket{0_{R}^{--}}^{(2)}
}
{
\bra{0_{R,--}}\sigma_{2}^{++}\left(w_{0}\right)\ket{0_{R}^{--}}^{\left(1\right)}\ket{0_{R}^{--}}^{\left(2\right)}} \,.
\label{amp-3}
\end{eqnarray}
From \eq{eq:ffdef-2} we observe that
\begin{eqnarray}
\mathcal{A}_{3}&=&\bra{0_{R,--}}d_{k}^{--}\sigma_{2}^{++}\left(w_{0}\right)d_{-m}^{(1)++}\ket{0_{R}^{--}}^{(1)} \ket{0_{R}^{--}}^{(2)}\cr
\cr
&=&\sum_{l\geq 1}f_{ml}^{F(1)+}\bra{0_{R,--}}d_{k}^{--}d_{-l}^{++}\ket{\boldsymbol{\chi}}\cr
&=&-C_{MN}f_{mk}^{F(1)+}
\end{eqnarray}
So we have
\begin{eqnarray}
f_{mk}^{F(1)+}=-\frac{\mathcal{A}_{3}}{\mathcal{A}_{1}}=-\frac{{}_{t}\bra{0_{NS}}d'^{--}_{k}d'^{(1)++}_{-m}\ket{0_{NS}}_{t}}{{}_{t}\langle 0_{NS}|0_{NS}\rangle_{t}}
\label{f_amp}
\end{eqnarray}
where on the RHS, the t-plane amplitude is after all spectral flows and coordinate shifts have been carried out for the copy 1 quantities $f^{F(1)\pm}$. For this calculation, we use the $t$ coordinate in which copy 1 is at the origin. When we calculate the copy 2 quantities $f^{F(2)\pm}$ we will use the $\hat{t}$ coordinate. 

We now compute the empty $t$ plane amplitude
\begin{eqnarray}
-\mathcal{A}_{3}&=&-{}_{t}\bra{0_{NS}}d'^{--}_{k}d'^{(1)++}_{-m}\ket{0_{NS}}_{t}
\end{eqnarray}
using the mode expansions in (\ref{d_t(i)_modes}) and (\ref{d_t_modes}). Let us first expand $d'^{--}_{k}$ around $t=\infty$. From (\ref{d_t_modes}) we have
\begin{eqnarray}
d'^{--}_{k}&=&\frac{1}{2\pi i}\oint\limits_{t=\infty}dt \,\psi^{--}(t)\left[t^{\frac{Mk}{M+N}}\left(t-a\right)^{\frac{Nk}{M+N}}\right].
\label{d_prime_f1}
\end{eqnarray}
We therefore expand
\begin{eqnarray}
t^{\frac{Mk}{M+N}}\left(t-a\right)^{\frac{Nk}{M+N}}&=&t^{k}\left(1-at^{-1}\right)^{\frac{Nk}{M+N}}=\sum_{p\geq 0}{}^{\frac{Nk}{M+N}}C_{p}(-a)^{p}t^{k-p} \,.
\end{eqnarray}
For our purposes, the sum is truncated by the requirement that $\tilde{d}_{k-p+\frac{1}{2}}^{--}$ be an annihilation operator.
So in terms of modes natural to the $t$-plane we find
\begin{eqnarray}
d'^{--}_{k}&\to&\sum_{p=0}^{k}{}^{\frac{Nk}{M+N}}C_{p}(-a)^{p}\tilde{d}^{--}_{k-p+\frac{1}{2}} \,.
\end{eqnarray}
Next, using (\ref{d_t(i)_modes}), we expand
\begin{eqnarray}
d'^{(1)++}_{-m}=\frac{\sqrt{M+N}}{2\pi i\sqrt{M}}\oint\limits_{t=0}dt\,\psi^{(1)++}(t)\left[t^{-m-1}\left(t-\tfrac{Ma}{M+N}\right)\left(t-a\right)^{-\frac{Nm}{M}-1}\right]
\label{d1_prime_f1}
\end{eqnarray}
We find
\begin{eqnarray}
&&t^{-m-1}\left(t-\tfrac{M}{M+N}a\right)\left(t-a\right)^{-\frac{Nm}{M}-1}\cr
&&\,\,\,\,\,\,\,\,\,\,\,\,\,\,\,=(-a)^{-\frac{Nm}{M}-1}\left(t^{-m}-\tfrac{M}{M+N}at^{-m-1}\right)\sum_{p'\geq 0}{}^{-\frac{Nm}{M}-1}C_{p'}(-a)^{-p'}t^{p'}\cr
&&\,\,\,\,\,\,\,\,\,\,\,\,\,\,\,=(-a)^{-\frac{Nm}{M}-1}\left[\sum_{p'\geq 0}{}^{-\frac{Nm}{M}-1}C_{p'}(-a)^{-p'}t^{p'-m}+\sum_{p'\geq 0}{}^{-\frac{Nm}{M}-1}C_{p'}\tfrac{M}{M+N}(-a)^{1-p'}t^{p'-m-1}\right]\nn
\end{eqnarray}
This gives
\begin{eqnarray}
d'^{(1)++}_{-m}&\to&\frac{\sqrt{M+N}}{\sqrt{M}}\sum_{p'=0}^{m-1}{}^{-\frac{Nm}{M}-1}C_{p'}(-a)^{-\frac{Nm}{M}-p'-1}\tilde{d}^{++}_{p'-m+\frac{1}{2}}\cr
&& {} + \frac{\sqrt{M}}{\sqrt{M+N}}\sum_{p'=0}^{m}{}^{-\frac{Nm}{M}-1}C_{p'}(-a)^{-\frac{Nm}{M}-p'}\tilde{d}^{++}_{p'-m-\frac{1}{2}}
\label{exp}
\end{eqnarray}
where again the upper limits on the sums are determined by the requirement that the operators on the RHS be creation operators. We now use the mode expansions to compute 
\begin{eqnarray}
-{}_{t}\bra{0_{NS}}d'^{++}_{k}d'^{(1)--}_{-m}\ket{0_{NS}}_{t}
\end{eqnarray} 
Since the expansion of $d'^{(1)++}_{-m}$ involves two separate terms, each involving a sum, we will separately calculate the contributions from these terms and add the resulting expressions to find the above amplitude. The contribution to the amplitude from the first term in the expansion of $d'^{(1)++}_{m}$ in (\ref{exp}) is
\begin{eqnarray}
-\sum_{p=0}^{k}\sum_{p'=0}^{m-1}\frac{\sqrt{M+N}}{\sqrt{M}}{}^{\frac{Nk}{M+N}}C_{p}{}^{-\frac{Nm}{M}-1}C_{p'}(-a)^{-\frac{Nm}{M}+(p-p')-1}{}_{t}\bra{0_{NS}}\tilde{d}^{--}_{k-p+\frac{1}{2}}\tilde{d}^{++}_{p'-m+\frac{1}{2}}\ket{0_{NS}}_{t}
\end{eqnarray}
where we have used the modes natural to the $t$ plane given in (\ref{nat_t_modes}). Using the anticommutation relations in (\ref{anti_comm}) we have the following constraints on the mode numbers
\begin{eqnarray}
k-p+\frac{1}{2}&=&-\left(p'-m+\frac{1}{2} \right)
\end{eqnarray}
This gives
\begin{eqnarray}
&&\frac{\sqrt{M+N}}{\sqrt{M}}(-a)^{k-\frac{m\left(M+N\right)}{M}}\sum_{p'=\max(m-k-1,0)}^{m-1}{}^{\frac{Nk}{M+N}}C_{k-m+p'+1}{}^{{}^{-\frac{Nm}{M}-1}}C_{p'}\,{}_{t}\langle 0_{NS}|0_{NS}\rangle_{t}\quad
\label{1st_term}
\end{eqnarray}
For the second term we find 
\begin{eqnarray}
-\frac{\sqrt{M}}{\sqrt{M+N}}\sum_{p=0}^{k}\sum_{p'=0}^{m}{}^{\frac{Nk}{M+N}}C_{p}{}^{-\frac{Nm}{M}-1}C_{p'}(-a)^{-\frac{Nm}{M}+(p-p')}{}_{t}\bra{0_{NS}}\tilde{d}^{--}_{k-p+\frac{1}{2}}\tilde{d}^{++}_{p'-m-\frac{1}{2}}\ket{0_{NS}}_{t} \,
\end{eqnarray}
which upon using the commutation relations becomes
\begin{eqnarray}
\frac{\sqrt{M}}{\sqrt{M+N}}(-a)^{k-\frac{m\left(M+N\right)}{M}}\sum_{p'=\max(m-k,0)}^{m}{}^{\frac{Nk}{M+N}}C_{k-m+p'}{}^{-\frac{Nm}{M}-1}C_{p'}\;{}_{t}\langle 0_{NS}|0_{NS}\rangle_{t} \,.
\label{2nd_term}
\end{eqnarray}
Adding together (\ref{1st_term}) and (\ref{2nd_term}),  and using (\ref{f_amp}), we find
\begin{eqnarray}
f_{mk}^{F(1)+}
&=&\frac{\sqrt{M+N}}{\sqrt{M}}(-a)^{k-\frac{m(M+N)}{M}}\sum_{p'=\max(m-k-1,0)}^{m-1}{}^{\frac{Nk}{M+N}}C_{k-m+p'+ 1}{}^{{}^{-\frac{Nm}{M}-1}}C_{p'}\cr
&&{}+\frac{\sqrt{M}}{\sqrt{M+N}}(-a)^{k-\frac{m\left(M+N\right)}{M}}\sum_{p'=\text{max}(m-k,0)}^{m}{}^{\frac{Nk}{M+N}}C_{k-m+p'}{}^{-\frac{Nm}{M}-1}C_{p'}
\label{f_sum}
\end{eqnarray}
Evaluating the sums, we obtain
\begin{eqnarray}
f_{mk}^{F(1)+}=\frac{(-1)^{m}\sqrt{M}\sin\left(\pi\frac{Mk}{M+N}\right)}{\pi\left(M+N\right)^{\frac{3}{2}}}\frac{(-a)^{k-\frac{m(M+N)}{M}}}{\frac{k}{M+N}-\frac{m}{M}}\frac{k}{m}\frac{\Gamma\left[\frac{Mk}{M+N}\right]\Gamma\left[\frac{Nk}{M+N}\right]}{\Gamma\left[k\right]}\frac{\Gamma\left[\frac{(M+N)m}{M}\right]}{\Gamma\left[m\right]\Gamma\left[\frac{Nm}{M}\right]}
\,. \quad
\label{f_F1+}
\end{eqnarray}
Note that in the above expression, when we have
\begin{eqnarray}
\frac{k}{M+N}=\frac{m}{M}
\end{eqnarray}
both numerator and denominator vanish, since
\begin{eqnarray}
\sin\left(\pi\tfrac{Mk}{M+N}\right)=\sin\left(\pi m\right)=0 \,.
\end{eqnarray}
Since the above expression for $f_{mk}^{F(1)+}$ is indeterminate in this situation, we return to the sum in (\ref{f_sum}), and take parameter values
\begin{eqnarray}
m~=~Mc \,,\qquad k&=&(M+N)c \,, \qquad c~=~\frac{j}{Y}
\end{eqnarray} 
where $j$ is a positive integer and $Y=\text{gcd}(M,N)$. Since $m<k$, we have 
\begin{eqnarray}
f_{mk}^{F(1)+}&=&\frac{\sqrt{M+N}}{\sqrt{M}}(-a)^{k-\frac{m(M+N)}{M}}\sum_{p'= 0}^{m-1}{}^{\frac{Nk}{M+N}}C_{k-m+p'+1}{}^{{}^{-\frac{Nm}{M}-1}}C_{p'}\cr
&& {}+\frac{\sqrt{M}}{\sqrt{M+N}}(-a)^{k-\frac{m\left(M+N\right)}{M}}\sum_{p'=0}^{m}{}^{\frac{Nk}{M+N}}C_{k-m+p'}{}^{-\frac{Nm}{M}-1}C_{p'}\cr
&=&\frac{\sqrt{M+N}}{\sqrt{M}}\sum_{p'= 0}^{Mc-1}{}^{Nc}C_{Nc+p'+1}{}^{-Nc-1}C_{p'}
+\frac{\sqrt{M}}{\sqrt{M+N}}\sum_{p'=0}^{Mc}{}^{Nc}C_{Nc+p'}{}^{-Nc-1}C_{p'}\cr
&=&\frac{\sqrt{M}}{\sqrt{M+N}} \,.
\label{s=q}
\end{eqnarray}

\subsection{Computing $f_{mk}^{F(1)-}$}

Next we compute $f^{F(1)-}$. The method is entirely analogous; this time, we take the amplitude in the numerator to be 
\begin{eqnarray}
\mathcal{A}_4&=&\bra{0_{R,--}}d_{k}^{++}\sigma_{2}^{+}\left(w_{0}\right)d'^{(1)--}_{-m}\ket{0_{R}^{--}}^{(1)}\ket{0_{R}^{--}}^{(2)}\cr
&=&-C_{MN}f_{mk}^{F(1)-} \,,
\end{eqnarray}
so we have 
\begin{eqnarray}
f_{mk}^{F(1)-}=-\frac{\mathcal{A}_{4}}{\mathcal{A}_{1}}=-\frac{{}_{t}\bra{0_{NS}}d'^{++}_{k}d'^{(1)--}_{-m}\ket{0_{NS}}_t}{{}_t\langle 0_{NS}|0_{NS}\rangle_{t}} \,
\label{f-amp}
\end{eqnarray}
where again on the RHS the amplitudes are in the empty $t$ plane after all spectral flows. 

Expanding the modes as before, we obtain
\begin{eqnarray}\label{f-sum}
f_{mk}^{F(1)-}&=&\frac{1}{\sqrt{M(M+N)}}(-a)^{k-\frac{(M+N)m}{M}}\\
&&\times\left[(M+N)\sum_{p'=\text{max}(m-k,0)}^{m-1}{}^{\frac{Nk}{M+N}-1}C_{k-m+p'}{}^{-\frac{Nm}{M}}C_{p'} \right. \cr
&&{}\left.
+M\sum_{p'=\text{max}(m-k+1,0)}^{m-1}{}^{\frac{Nk}{M+N}-1}C_{k-m+p'-1}{}^{-\frac{Nm}{M}}C_{p'}\right] . \nonumber 
\end{eqnarray}
Evaluating the sums, we find
\begin{eqnarray}
f_{mk}^{F(1)-}=\frac{(-1)^{m}\sin\left(\pi\frac{Mk}{M+N}\right)}{\pi\sqrt{M\left(M+N\right)}}\frac{(-a)^{k-\frac{m(M+N)}{M}}}{\frac{k}{M+N}-\frac{m}{M}}\frac{\Gamma\left[\frac{Mk}{M+N}\right]\Gamma\left[\frac{Nk}{M+N}\right]}{\Gamma\left[k\right]}\frac{\Gamma\left[\frac{(M+N)m}{M}\right]}{\Gamma\left[m\right]\Gamma\left[\frac{Nm}{M}\right]} .
\label{final_f1_-}
\end{eqnarray}
As before, for the special case of
\begin{eqnarray}
\frac{k}{M+N}=\frac{m}{M}
\end{eqnarray} we have an indeterminate expression for $f_{mk}^{F(1)-}$ and using (\ref{f-sum}) we find 
\begin{eqnarray}
f_{mk}^{F(1)-}&=&\frac{1}{\sqrt{M(M+N)}}\!
\left[(M+N)\sum_{p'=0}^{Mc-1}{}^{Nc-1}C_{Nc+p'}{}^{-Nc}C_{p'}+M\sum_{p'=0}^{Mc-1}{}^{Nc-1}C_{Nc+p'-1}{}^{-Nc}C_{p'}\right]\cr
&=&\frac{\sqrt{M}}{\sqrt{M+N}} \,.
\label{s=q_2-}
\end{eqnarray}

\subsection{Computing $f_{mk}^{F(2)+}$ and $f_{mk}^{F(2)-}$}
For an initial excitation on copy 2, we calculate $f_{mk}^{F(2)+}$. Similar to the calculation of $f_{mk}^{F(1)+}$, we consider the amplitude
\begin{eqnarray}
\mathcal{A}_{5}&=&\bra{0_{R,--}}d_{k}^{--}\sigma_{2}^{++}\left(w_{0}\right)d_{-m}^{(2)++}\ket{0_{R}^{--}}^{(1)} \ket{0_{R}^{--}}^{(2)}\cr
\cr
&=&\sum_{l\geq 1}f_{ml}^{F(2)+}\bra{0_{R,--}}d_{k}^{--}d_{-l}^{++}\ket{\boldsymbol{\chi}}\cr
&=&-C_{MN}f_{mk}^{F(2)+}
\end{eqnarray}
So we have
\begin{eqnarray}
f_{mk}^{F(2)+}=-\frac{\mathcal{A}_{4}}{\mathcal{A}_{1}}=-\frac{{}_{t}\bra{0_{NS}}d'^{--}_{k}d'^{(2)++}_{-m}\ket{0_{NS}}_{t}}{{}_{t}\langle 0_{NS}|0_{NS}\rangle_{t}} \,.
\label{f_amp_2}
\end{eqnarray}
Let us now compute the empty $\hat{t}$ plane amplitude
\begin{eqnarray}
&&{}_{\hat{t}}\bra{0_{NS}}d'^{--}_{k}d'^{(2)++}_{-m}\ket{0_{NS}}_{\hat{t}}\cr
&&\quad={}_{\hat{t}}\bra{0_{NS}}\frac{1}{2\pi i}\oint\limits_{\hat{t}=\infty}d\hat{t}\psi^{--}(\hat{t})\left[(\hat{t}+a)^{\frac{Mk}{M+N}}\hat{t}^{\frac{Nk}{M+N}}\right]\cr
&&\quad\quad\times\frac{\sqrt{M+N}}{\sqrt{N}}\frac{1}{2\pi i}\oint\limits_{\hat{t}=0}d\hat{t}\psi^{++}(\hat{t})\left[(\hat{t}+a)^{-\frac{Mm}{N}-1}\left(\hat{t}+\frac{Na}{M+N}\right)\hat{t}^{-m-1}\right]\ket{0_{NS}}_{\hat{t}}\qquad
\label{f_2_+}
\end{eqnarray}
Looking at (\ref{f_2_+}), (\ref{d_prime_f1}) and (\ref{d1_prime_f1}), we notice that if we make following interchanges
\begin{eqnarray}
M\leftrightarrow N,\quad a\leftrightarrow -a
\label{interchanges}
\end{eqnarray}
then
\begin{eqnarray} 
f^{F(2)+}_{mk}\leftrightarrow f^{F(1)+}_{mk}
\end{eqnarray}
Therefore we can use the results obtained for the expression $f^{F(1)+}_{mk}$ and make the interchanges given in (\ref{interchanges}) to obtain the expression for $f^{F(2)+}_{mk}$. The same applies for 
\begin{eqnarray} 
f^{F(2)-}_{mk}\leftrightarrow f^{F(1)-}_{mk} \,.
\end{eqnarray}
Then from (\ref{f_F1+}), (\ref{interchanges}) we find that for the case 
$
\frac{k}{M+N}\neq \frac{m}{N}
$
we have
\begin{eqnarray}
f_{mk}^{F(2)+} &=& \frac{(-1)^{m}\sqrt{N}\sin\left(\pi\frac{Nk}{M+N}\right)}{\pi\left(M+N\right)^{\frac{3}{2}}} \, \frac{a^{k-\frac{m(M+N)}{N}}}{\frac{k}{M+N}-\frac{m}{N}} 
\,\frac{k}{m} \,
\frac{\Gamma\left[\frac{Mk}{M+N}\right]\Gamma\left[\frac{Nk}{M+N}\right]}{\Gamma\left[k\right]}
\,
\frac{\Gamma\left[\frac{(M+N)m}{N}\right]}{\Gamma\left[m\right]\Gamma\left[\frac{Mm}{N}\right]}
\label{f_F2+} \qquad \\
f_{mk}^{F(2)-}&=&
\frac{(-1)^{m}\sin\left(\pi\frac{Nk}{M+N}\right)}{\pi\sqrt{N\left(M+N\right)}}
 \, \frac{a^{k-\frac{m(M+N)}{N}}}{\frac{k}{M+N}-\frac{m}{N}} \, 
\frac{\Gamma\left[\frac{Mk}{M+N}\right]\Gamma\left[\frac{Nk}{M+N}\right]}{\Gamma\left[k\right]}
\, \frac{\Gamma\left[\frac{(M+N)m}{N}\right]}{\Gamma\left[m\right]\Gamma\left[\frac{Mm}{N}\right]}
\end{eqnarray}
and for the case of
$
\frac{k}{M+N}=\frac{m}{N}
$
we have
\begin{eqnarray}
f_{mk}^{F(2)+} ~=~ f_{mk}^{F(2)-} &=& \frac{\sqrt{N}}{\sqrt{M+N}}\,.
\eea

\subsection{Expressing $f_{mk}^{F(i)\pm}$ in final form}

To express $f_{mk}^{F(i)\pm}$ in final form, we now make analogous changes of notation as done for $\gamma_{kl}^{F}$.
\begin{enumerate}
\item[(i)] Using \eq{az}, we replace the parameter $a$ by $z_{0}=e^{w_0}$.
\item[(ii)] For component string (1) we use fractional modes
\begin{eqnarray}
q=\frac{m}{M},&&s=\frac{k}{M+N}
\end{eqnarray}
and for component string  (2) we use fractional modes
\begin{eqnarray}
r=\frac{m}{N},&&s=\frac{k}{M+N} \,.
\end{eqnarray}
When expressing our result for component string (1) using indices $q$ and $s$, we write
\begin{eqnarray}
f_{mk}^{F(1)\pm}\to\tilde{f}_{qs}^{F(1)\pm}
\end{eqnarray} 
and for component string (2) using indices $r$ and $s$, we write
\begin{eqnarray}
f_{mk}^{F(2)\pm}\to\tilde{f}_{rs}^{F(2)\pm} \,.
\end{eqnarray}
\item[(iii)] We use the shorthand $\mu_{s}=(1-e^{2\pi i Ms})$. In doing this we note that
\begin{eqnarray}
\sin\left(\pi Ms\right)
=\frac{i}{2}e^{-i\pi Ms}\mu_s \,.
\end{eqnarray}
\end{enumerate}
With these changes in notation, for component string (1) with $s\neq q$ we have
\begin{eqnarray}
\tilde{f}_{qs}^{F(1)+}&=&\frac{i}{2\pi}z_{0}^{s-q}\frac{\mu_{s}}{\sqrt{M(M+N)}}\frac{1}{s-q}
 \, \frac{s}{q}\left(\frac{(M+N)^{M+N}}{M^{M}N^{N}}\right)^{s-q}
\frac{\Gamma\left[(M+N)q\right]}{\Gamma\left[Mq\right]\Gamma\left[Nq\right]}
\frac{\Gamma\left[Ms\right]\Gamma\left[Ns\right]}{\Gamma\left[(M+N)s\right]}
\cr
\tilde{f}_{qs}^{F(1)-}&=&\frac{i}{2\pi}z_{0}^{s-q}\frac{\mu_{s}}{\sqrt{M(M+N)}}\frac{1}{s-q}
\left(\frac{(M+N)^{M+N}}{M^{M}N^{N}}\right)^{s-q}
\frac{\Gamma\left[(M+N)q\right]}{\Gamma\left[Mq\right]\Gamma\left[Nq\right]}
\frac{\Gamma\left[Ms\right]\Gamma\left[Ns\right]}{\Gamma\left[(M+N)s\right]}
\nn
\label{f+}
\end{eqnarray}
For component string (2) with $s\neq r$ we obtain
\begin{eqnarray}
\tilde{f}_{rs}^{F(2)+}&=&-\frac{i}{2\pi}z_{0}^{s-r}\frac{\mu_{s}}{\sqrt{N(M+N)}}\frac{1}{s-r}
 \, \frac{s}{r}\left(\frac{(M+N)^{M+N}}{M^{M}N^{N}}\right)^{s-r}
\frac{\Gamma\left[(M+N)r\right]}{\Gamma\left[Mr\right]\Gamma\left[Nr\right]}
\frac{\Gamma\left[Ms\right]\Gamma\left[Ns\right]}{\Gamma\left[(M+N)s\right]}
\cr
\tilde{f}_{rs}^{F(2)-}&=&-\frac{i}{2\pi}z_{0}^{s-r}\frac{\mu_{s}}{\sqrt{N(M+N)}}\frac{1}{s-r}
\left(\frac{(M+N)^{M+N}}{M^{M}N^{N}}\right)^{s-r}
\frac{\Gamma\left[(M+N)r\right]}{\Gamma\left[Mr\right]\Gamma\left[Nr\right]}
\frac{\Gamma\left[Ms\right]\Gamma\left[Ns\right]}{\Gamma\left[(M+N)s\right]}
\nn
\label{f2+}
\end{eqnarray}
For $M=N=1$, given the normalizations chosen, the quantities $\frac{1}{\sqrt{2}}f^{F(i)+}$ in this paper should agree with the  $f^{F(i)+}$ computed in \cite{\acmthree}. One can check that this is indeed the case.

\subsection{Applying the supercharge for an initial fermionic excitation}

In this section we include the supercharge, in order to obtain the full effect of the deformation operator on an initial fermionic excitation. We  introduce the notation
\begin{eqnarray}
\ket{\Psi^{F(i)\pm}}&=&\hat{O}_{\dot{A}}(w_0)d^{(i)\pm B}_{-m}\ket{0_{R}^{--}}^{(1)} \ket{0_{R}^{--}}^{(2)} \,.
\end{eqnarray}
As before we introduce separate notation for the holomorphic and antiholomorphic parts, 
\begin{eqnarray}
\ket{\Psi^{F(i)\pm}}&=& \ket{\psi^{F(i)\pm}} \ket{\bar\psi^{F(i)\pm}} \,.
\end{eqnarray}

Since we now have have an initial excitation, we must compute the contribution of the supercharge operator above and below the twist insertion. Writing only the holomorphic parts, recall that
\be
\hat{O}_{\dot A} = G^{-}_{\dot A,0}\s_2^+ - \s_2^+ \left ( G^{(1)-}_{\dot A,0} + G^{(2)-}_{\dot A,0}\right )
\ee
where from \eq{G0_below-1} we recall
\begin{eqnarray}
-G_{\dot{A},0}^{(1)-}&=&-\frac{1}{2 \pi i}\int\limits_{w=\tau_{0}-\epsilon}^{\tau_{0}-\epsilon+2\pi iM}G_{\dot{A}}^{(1)-}(w)dw~=~-\frac{i}{\sqrt{M}}\sum_{n}d^{(1)-A}_{n}\alpha_{A\dot{A},-n}^{(1)}
\end{eqnarray} 
and similarly for component string (2).
There four cases to compute: $\ket{\psi^{F(1)+}}$, $\ket{\psi^{F(1)-}}$, 
$\ket{\psi^{F(2)+}}$, and $\ket{\psi^{F(2)-}}$. 

Before proceeding, we record the expression for $f_{mk}^{B(1)}$ calculated in \cite{\chmtone}:
\begin{eqnarray}
f_{mk}^{B(1)}=\frac{(-1)^{m}\sin\left(\pi\frac{Mk}{M+N}\right)}{\pi\left(M+N\right)}\frac{(-a)^{k-\frac{m(M+N)}{M}}}{\frac{k}{M+N}-\frac{m}{M}}\frac{\Gamma\left[\frac{Mk}{M+N}\right]\Gamma\left[\frac{Nk}{M+N}\right]}{\Gamma\left[k\right]}\frac{\Gamma\left[\frac{(M+N)m}{M}\right]}{\Gamma\left[m\right]\Gamma\left[\frac{Nm}{M}\right]} \,.
\label{eq:fb}
\end{eqnarray}

\subsubsection{$\ket{\psi^{F(1)+}}$}

Let us begin with $\ket{\psi^{F(1)+}}$.
Our final state is given by (we write only the holomorphic parts)
\begin{eqnarray}
\ket{\psi^{F(1)+}}=\left(G_{\dot{A},0}^{-}\sigma_{2}^{+}(w_{0})d^{(1)+B}_{-m}-\sigma_{2}^{+}G_{\dot{A},0}^{(1)-}d^{(1)+B}_{-m}\right)\ket{0_{R}^{-}}^{(1)} \ket{0_{R}^{-}}^{(2)}
\end{eqnarray}
Using (\ref{eq:ffdef}) the first term is given by
\begin{eqnarray}
G_{\dot{A},0}^{-}\sum_{k\geq 1}f_{mk}^{F(1)+}d_{-k}^{+ B}\ket{\chi}&=&\frac{i\epsilon^{AB}}{\sqrt{M+N}}\sum_{k\geq 1}f_{mk}^{F(1)+}\alpha_{A\dot{A},-k}\ket{\chi}+\sum_{k\geq 1}f_{mk}^{F(1)+}d^{+B}_{-k}\ket{\psi} \nn
\end{eqnarray}
where we recall the notation $\ket{\psi}=G_{\dot{A},0}^{-}\ket{\chi}$. $\ket{\psi}$ is given explicitly in (\ref{eq:psians}).

For the second term, using $G_{\dot{A},0}^{(1)-}\ket{0_{R}^{-}}^{(1)}=0$ we have
\begin{eqnarray}
\sigma_{2}^{+}(w_{0})G_{\dot{A},0}^{(1)-}d^{(1)+B}_{-m}\ket{0_{R}^{-}}^{(1)} \ket{0_{R}^{-}}^{(2)}&=&\sigma_{2}^{+}(w_{0})\lbrace G_{\dot{A},0}^{(1)-},d_{-m}^{(1)+B}\rbrace\ket{0_{R}^{-}}^{(1)} \ket{0_{R}^{-}}^{(2)}\\
&=&\frac{i \epsilon^{AB}}{\sqrt{M}}\sum_{k\geq 1}f_{mk}^{B(1)}\alpha_{A\dot{A},-k}\ket{\chi} \,.
\end{eqnarray}
Combining both terms we obtain
\begin{eqnarray}
\ket{\psi^{F(1)+}}&=&i\epsilon^{AB}
\sum_{k\geq 1}
\left(\frac{f_{mk}^{F(1)+}}{\sqrt{M+N}}-\frac{f_{mk}^{B(1)}}{\sqrt{M}}\right)
\alpha_{A\dot{A},-k}
\ket{\chi}+\sum_{k\geq 1}f_{mk}^{F(1)+}d_{-k}^{+B}\ket{\psi} 
\label{final_f(1)+}
\eea
where we note that
\bea
&&\frac{f_{mk}^{F(1)+}}{\sqrt{M+N}}-\frac{f_{mk}^{B(1)}}{\sqrt{M}} \cr
&&{} \quad =
\frac1\pi \frac{\sqrt{M}}{M+N} \frac{(-1)^{m}}{m}  \sin\left(\pi\frac{Mk}{M+N}\right)
(-a)^{k-\frac{m(M+N)}{M}}
\frac{\Gamma\left[\frac{Mk}{M+N}\right]\Gamma\left[\frac{Nk}{M+N}\right]}{\Gamma\left[k\right]}\frac{\Gamma\left[\frac{(M+N)m}{M}\right]}{\Gamma\left[m\right]\Gamma\left[\frac{Nm}{M}\right]} \,.  \cr &&
\label{final_f(1)+2}
\end{eqnarray}

\subsubsection{$\ket{\psi^{F(1)-}}$}

Here our final state is given by 
\begin{eqnarray}
\ket{\psi^{F(1)-}}=
\left(G_{\dot{A},0}^{-}\sigma_{2}^{+}(w_{0})d^{(1)-B}_{-m}
-\sigma_{2}^{+}G_{\dot{A},0}^{(1)-}d^{(1)-B}_{-m}  \right)
\ket{0_{R}^{-}}^{(1)} \ket{0_{R}^{-}}^{(2)}
\label{psi_f(1)-}
\end{eqnarray}
For the first term we find
\begin{eqnarray}
G_{\dot{A},0}^{-}\sigma_{2}^{+}d_{-m}^{(1)-B}\ket{0_{R}^{-}}^{(1)} \ket{0_{R}^{-}}^{(2)}&=&G_{\dot{A},0}^{-}\sum_{k\geq 1}f_{mk}^{F(1)-}d_{-k}^{-B}\ket{\chi}
~=~\sum_{k\geq 1}f_{mk}^{F(1)-}d_{-k}^{-B}\ket{\psi}\qquad
\end{eqnarray}
and for the second term, using $G_{\dot{A},0}^{(1)-}\ket{0_{R}^{-}}^{(1)}=0$ we have
\begin{eqnarray}
\sigma_{2}^{+}G_{\dot{A},0}^{(1)-}d_{-m}^{(1)-B}\ket{0_{R}^{-}}^{(1)} \ket{0_{R}^{-}}^{(2)}=0 \,.
\end{eqnarray}
Therefore (\ref{psi_f(1)-}) becomes 
\begin{eqnarray}
\ket{\psi^{F(1)-}}&=&-\sum_{k\geq 1}f_{mk}^{F(1)-}d_{-k}^{-B}\ket{\psi} \,.
\label{final_f(1)-}
\end{eqnarray}

\subsubsection{$\ket{\psi^{F(2)+}}$ and $\ket{\psi^{F(2)-}}$}

The states $\ket{\psi^{F(2)+}}$ and $\ket{\psi^{F(2)+}}$ may be computed by modifying (\ref{final_f(1)+}) and (\ref{final_f(1)-}), respectively, by $M\leftrightarrow N$, $a\rightarrow -a$, and $(1) \rightarrow (2)$.

\newpage

\subsection{Applying the supercharge for an initial bosonic excitation}

We now compute the state produced when the deformation operator acts on an initial bosonic excitation. 
In the same way done for an initial fermionic excitation, we introduce the notation
\begin{eqnarray}
\ket{\Psi^{B(i)}} &=& 
\hat{O}_{\dot{A}}(w_0)\, \a^{(i)}_{B\dot B,-m}\ket{0_{R}^{--}}^{(1)} \ket{0_{R}^{--}}^{(2)} ~=~ 
\ket{\psi^{B(i)}}\ket{\bar\psi^{B(i)}} \,.
\eea

\subsubsection{$\ket{\psi^{B(1)}}$}

We first compute the state $\ket{\psi^{B(1)}}$. We have
\begin{eqnarray}
\ket{\psi^{B(1)}} 
&=&
\left(G_{\dot{A},0}^{-}\sigma_{2}^{+}(w_{0})\alpha^{(1)}_{B\dot B,-m}-\sigma_{2}^{+}G_{\dot{A},0}^{(1)-}\alpha^{(1)}_{B\dot B,-m}\right)\ket{0_{R}^{-}}^{(1)} \ket{0_{R}^{-}}^{(2)}
\end{eqnarray}
The first term becomes
\begin{eqnarray}
G_{\dot{A},0}^{-}\sum_{k\geq 1}f_{mk}^{B(1)}\alpha_{B\dot{B},-k}\ket{\chi}&=&i\epsilon_{AB}\epsilon_{\dot{A}\dot{B}}\sum_{k\geq 1}\left(\frac{k \, f_{mk}^{B(1)}}{\sqrt{M+N}}\right)d_{-k}^{-A}\ket{\chi}+\sum_{k\geq 1}f_{mk}^{B(1)}\alpha_{B\dot{B},-k}\ket{\psi}\nonumber
\end{eqnarray}
For the second term, we find
\begin{eqnarray}
\sigma_{2}^{+}G_{\dot{A},0}^{(1)-}\alpha^{(1)}_{B\dot{B},-m}\ket{0_{R}^{-}}^{(1)} \ket{0_{R}^{-}}^{(2)}&=&i\epsilon_{AB}\epsilon_{\dot{A}\dot{B}}\frac{m}{\sqrt{M}}\sigma_{2}^{+}d_{-m}^{(1)-A}\ket{0_{R}^{-}}^{(1)} \ket{0_{R}^{-}}^{(2)}\cr
&=&i\epsilon_{AB}\epsilon_{\dot{A}\dot{B}}\sum_{k\geq 1}
\left(
\frac{m \, f_{mk}^{F(1)-}}{\sqrt{M}} \right)
d_{-k}^{-A}\ket{\chi}
\end{eqnarray}
Combining both terms, we obtain
\begin{eqnarray}
\ket{\psi^{B(1)}}&=&i\epsilon_{AB}\epsilon_{\dot{A}\dot{B}}
\sum_{k\geq 1}
\left(\frac{k \, f_{mk}^{B(1)}}{\sqrt{M+N}}-\frac{m \, f_{mk}^{F(1)-}}{\sqrt{M}}\right)
d_{-k}^{-A}\ket{\chi}+\sum_{k\geq 1}f_{mk}^{B(1)}\alpha_{B\dot{B},-k}\ket{\psi} \quad
\label{final_f(1)}
\eea
where we note that\footnote{To compare to \cite{\acmthree} in the limit of $ M = N = 1 $, one should take into account the different conventions on fermion modes. Note also that there is a typo in the last term of equation (7.5) of \cite{\acmthree}; this term should resemble the last term in \eq{final_f(1)} above.}
\bea
\frac{k \, f_{mk}^{B(1)}}{\sqrt{M+N}}-\frac{m \, f_{mk}^{F(1)-}}{\sqrt{M}}
 &=&
\frac{(-1)^{m}\sin\left(\pi\frac{Mk}{M+N}\right)}{\pi\sqrt{M+N}}
(-a)^{k-\frac{m(M+N)}{M}}
\frac{\Gamma\left[\frac{Mk}{M+N}\right]\Gamma\left[\frac{Nk}{M+N}\right]}{\Gamma\left[k\right]}
\frac{\Gamma\left[\frac{(M+N)m}{M}\right]}{\Gamma\left[m\right]\Gamma\left[\frac{Nm}{M}\right]}.
\cr &&
\label{final_f(1)-2}
\end{eqnarray}

\subsubsection{$\ket{\psi^{B(1)}}$}

To find $\ket{\psi^{B(2)}}$ one simply modifies (\ref{final_f(1)}) by $M\leftrightarrow N$, $a\rightarrow -a$, and $(1) \rightarrow (2)$. 

\subsection{Summary of Results}
For convenient reference, here we record the results for ${\gamma}^{F}$, $C_{MN}$ and $f^{(i)\pm}$. For completeness we include also the bosonic quantities $\g^B$, $f^B$ computed in \cite{\chmtone}.
\begin{eqnarray}
\t\gamma^B_{ss'}&=&{z_0^{s+s'}\over 4\pi^2}\,\mu_s\mu_{s'}\,
{1 \over  s+s'}\,{MN\over (M+N)^3} 
\left ({(M+N)^{M+N}\over M^MN^N}\right )^{s+s'} 
{\Gamma[{Ms}]\Gamma[{Ns}]\over \Gamma[(M+N)s]}~{\Gamma[{Ms'}]\Gamma[{Ns'}]\over \Gamma[(M+N)s']}\cr
\t\gamma_{ss'}^{F}&=&-\frac{z_{0}^{s+s'}}{4\pi^{2}}\mu_{s}\mu_{s'}\frac{s}{s+s'}\frac{MN}{(M+N)^{2}}\left(\frac{(M+N)^{M+N}}{M^{M}N^{N}}\right)^{s+s'}\frac{\Gamma\left[Ms\right]\Gamma\left[Ns\right]}{\Gamma\left[(M+N)s\right]}\frac{\Gamma\left[Ms'\right]\Gamma\left[Ns'\right]}{\Gamma\left[(M+N)s'\right]}\cr
C_{MN}&=&\frac{M+N}{2MN}
\end{eqnarray}
\bea
\t f^{B(1)}_{qs} & = & \begin{cases}
{M \over M+N} & q = s \\
{i\over 2\pi}z_0^{s-q}
{\mu_s \over s-q}\,
{1\over  (M+N)}\left( {(M+N)^{M+N}\over M^MN^N}\right)^{(s-q)}
{\Gamma[(M+N)q]\over \Gamma[Mq]\Gamma[Nq]}\,
{\Gamma[Ms]\Gamma[Ns]\over \Gamma[(M+N)s]}
\qquad\quad & q \neq s 
\end{cases}\qquad\quad \cr
\t f^{B(2)}_{rs} & = & \begin{cases}
{N \over M+N} & r = s \\
-{i\over 2\pi}z_0^{s-r}
{\mu_s\over s-r}\,
{1\over  (M+N)}\left( {(M+N)^{M+N}\over M^MN^N}\right)^{(s-r)}
{\Gamma[(M+N)r]\over \Gamma[Mr]\Gamma[Nr]}\,
 {\Gamma[Ms]\Gamma[Ns]\over \Gamma[(M+N)s]} \qquad\quad & r \neq s
\end{cases}\qquad\quad
\eea

\bea
\tilde{f}_{qs}^{F(1)+}&=&\begin{cases}
\frac{\sqrt{M}}{\sqrt{M+N}}& q=s\\
&\\
\frac{i}{2\pi}z_{0}^{s-q}\frac{\mu_{s}}{\sqrt{M(M+N)}}\frac{1}{s-q}
 \, \frac{s}{q}\left(\frac{(M+N)^{M+N}}{M^{M}N^{N}}\right)^{s-q}
\frac{\Gamma\left[(M+N)q\right]}{\Gamma\left[Mq\right]\Gamma\left[Nq\right]}
\frac{\Gamma\left[Ms\right]\Gamma\left[Ns\right]}{\Gamma\left[(M+N)s\right]}
 & q\neq s
\end{cases} \cr
\tilde{f}_{qs}^{F(1)-}&=&
\begin{cases}\frac{\sqrt{M}}{\sqrt{M+N}}& q=s\\
&\\
\frac{i}{2\pi}z_{0}^{s-q}\frac{\mu_{s}}{\sqrt{M(M+N)}}\frac{1}{s-q}
\left(\frac{(M+N)^{M+N}}{M^{M}N^{N}}\right)^{s-q}
\frac{\Gamma\left[(M+N)q\right]}{\Gamma\left[Mq\right]\Gamma\left[Nq\right]}
\frac{\Gamma\left[Ms\right]\Gamma\left[Ns\right]}{\Gamma\left[(M+N)s\right]}
&q\neq s
\end{cases}
\end{eqnarray}

\bea
\tilde{f}_{rs}^{F(2)+}&=&\begin{cases}
\frac{\sqrt{N}}{\sqrt{M+N}}& r=s\\
&\\
-\frac{i}{2\pi}z_{0}^{s-r}\frac{\mu_{s}}{\sqrt{N(M+N)}}\frac{1}{s-r}
 \, \frac{s}{r}\left(\frac{(M+N)^{M+N}}{M^{M}N^{N}}\right)^{s-r}
\frac{\Gamma\left[(M+N)r\right]}{\Gamma\left[Mr\right]\Gamma\left[Nr\right]}
\frac{\Gamma\left[Ms\right]\Gamma\left[Ns\right]}{\Gamma\left[(M+N)s\right]}
 & r\neq s
\end{cases} \cr
\tilde{f}_{rs}^{F(2)-}&=&
\begin{cases}\frac{\sqrt{N}}{\sqrt{M+N}}& r=s\\
&\\
-\frac{i}{2\pi}z_{0}^{s-r}\frac{\mu_{s}}{\sqrt{N(M+N)}}\frac{1}{s-r}
\left(\frac{(M+N)^{M+N}}{M^{M}N^{N}}\right)^{s-r}
\frac{\Gamma\left[(M+N)r\right]}{\Gamma\left[Mr\right]\Gamma\left[Nr\right]}
\frac{\Gamma\left[Ms\right]\Gamma\left[Ns\right]}{\Gamma\left[(M+N)s\right]}
&r\neq s
\end{cases}
\end{eqnarray}

Since the expressions for the final states $\ket{\psi^{F(i)\pm}}$ and $\ket{\psi^{B(i)}}$ are somewhat unwieldy, we will not repeat their expressions here, and we instead refer the reader to equations \eq{final_f(1)+}, \eq{final_f(1)-}, \eq{final_f(1)}.

\section{Continuum Limit} \label{sec:cont}

Thus far, our computations have been exact, and the results are somewhat involved. To study black hole physics however, one would like to take the limit of large $N_{1}N_{5}$. In this limit, typical component strings have parametrically large winding numbers, and typical mode numbers $q$, $r$, $s$ are much larger than the spacing of modes on the respective component string. We denote this limit by the `continuum limit'. Large mode numbers correspond to short wavelengths, which are not sensitive to the finite length of the component string. In this limit, the expressions for $\gamma^{F}$ and $f^{F(i)\pm}$ simplify considerably, as we now show.

\subsection{Continuum Limit for $\gamma_{kl}^{F}$}

We start by taking the continuum limit of $\gamma^{F}$. From \eq{gen_gamma} we have 
\begin{eqnarray}
\tilde{\gamma}_{ss'}^{F}=-\frac{1}{4\pi^{2}}z_{0}^{s+s'}\mu_{s}\mu_{s'}\frac{s}{s+s'}\frac{MN}{(M+N)^{2}}\left(\frac{(M+N)^{M+N}}{M^{M}N^{N}}\right)^{s+s'}\frac{\Gamma\left[Ms\right]\Gamma\left[Ns\right]}{\Gamma\left[(M+N)s\right]}\frac{\Gamma\left[Ms'\right]\Gamma\left[Ns'\right]}{\Gamma\left[(M+N)s'\right]}\cr
\label{final_gamma}
\end{eqnarray}
We wish to find the approximation to this expression when 
\begin{eqnarray}
s >> \frac{1}{M+N},\quad s>>\frac{1}{M},\quad s>>\frac{1}{N}
\end{eqnarray}
and likewise for $s'$.
We use Stirling's formula, 
\begin{eqnarray}
\Gamma[x]\sim\sqrt{\frac{2\pi}{x}}\left(\frac{x}{e}\right)^{x}
\end{eqnarray}
for $x>>1$.
Using this, we get
\begin{eqnarray}
\Gamma[Ms]&\sim&\sqrt{\frac{2\pi}{Ms}}\left(\frac{Ms}{e}\right)^{Ms} \,. 
\end{eqnarray}
For the Gamma function terms of (\ref{final_gamma}) with variable $s$, we then have
\begin{eqnarray}
\frac{\Gamma\left[Ms\right]\Gamma\left[Ns\right]}
{\Gamma\left[(M+N)s\right]}
\approx
\sqrt{\frac{M+N}{MN}}\sqrt{\frac{2\pi}{s}}\left(\frac{M^{M}N^{N}}{(M+N)^{M+N}}\right)^{s}
\label{cont_limit}
\end{eqnarray}
and likewise for the terms with variable $s'$.

Inserting these approximations in (\ref{final_gamma}), we find
\begin{eqnarray}
\tilde{\gamma}_{ss'}^{F}&\approx&-\frac{1}{2\pi}z_{0}^{s+s'}\frac{\mu_{s}\mu_{s'}}{(M+N)}\frac{1}{s+s'}\sqrt{\frac{s}{s'}}\,.
\end{eqnarray}
Let us also obtain the continuum limit for $f^{F(i)\pm}$. We need only consider the cases where $s\neq q$ for $f_{qs}^{F(1)\pm}$ and $s\neq r$ for $f_{qs}^{F(2)\pm}$ because for the cases $s=q$ and $s=r$ the continuum limit expression is identical to the exact expression.

For $f^{F(1)+}$, from \eq{f+} we have the exact expression
\begin{eqnarray}
\tilde{f}_{qs}^{F(1)+}&=&\frac{i}{2\pi}z_{0}^{s-q}
\frac{\mu_{s}}{\sqrt{M(M+N)}}\frac{1}{s-q}
 \, \frac{s}{q}\left(\frac{(M+N)^{M+N}}{M^{M}N^{N}}\right)^{s-q}\frac{\Gamma\left[Ms\right]\Gamma\left[Ns\right]}{\Gamma\left[(M+N)s\right]}\frac{\Gamma\left[(M+N)q\right]}{\Gamma\left[Mq\right]\Gamma\left[Nq\right]}\cr &&
\end{eqnarray}
thus we obtain
\begin{eqnarray}
\tilde{f}_{qs}^{F(1)+}\approx\frac{i}{2\pi}z_{0}^{s-q}
\frac{\mu_{s}}{\sqrt{M(M+N)}}\frac{1}{s-q}
\sqrt{\frac{s}{q}} \,. \qquad
\end{eqnarray}
Similarly, we obtain
\begin{eqnarray}
\tilde{f}_{qs}^{F(1)-}\approx\frac{i}{2\pi}z_{0}^{s-q}
\frac{\mu_{s}}{\sqrt{M(M+N)}}\frac{1}{s-q}
\sqrt{\frac{q}{s}}
\end{eqnarray}
and the component string  (2) quantities can be found in the same way.

\section{Discussion} \label{sec:disc}

In this paper we have studied the effect of the deformation operator when it joins together two component strings of length $ M, N $ into a single component string of winding $ M+N$. We computed the final state produced from starting with a R-R vacuum state on the original component strings, and also in the case of an initial bosonic or fermionic excitation. 

Our results generalize those recently reported in \cite{\chmtone}, which studied bosonic fields only,  in two ways. Firstly,  we have extended the analysis to the fermionic fields. Secondly, we have computed the overall prefactor on the final state. The calculation of this prefactor is quite nontrivial, but the answer is compact: $C_{MN} = (M+N)/(2MN) $. If one considers the special case $ M=N $, this becomes $C_{MM} = 1/M$, which agrees with the fact that the twist operator $\s_2^{++}$ responsible for this coefficient has weight $(1/2,1/2)$.

It was noted in \cite{\chmtone} that the bosonic quantities $\gamma^B$, $f^{B(i)}$ have an interesting structure involving Gamma functions, which ensures that $\gamma^B$ multiplies only creation operators and that $f^{B(i)}$ vanishes unless there is a creation operator in the initial state, and a creation operator in the final state. As one would expect, we find that these characteristics are shared by the fermionic quantities $\gamma^F$, $f^{F(i)}$.

In addition, the fermionic quantities $\gamma^F$, $f^{F(i)}$ share the property of `almost factorization' observed for the bosonic quantities $\gamma^B$, $f^{B(i)}$ in~\cite{\chmtone}. For $\gamma^F_{ss'}$, the only part which does not factorize into a product of terms corresponding to $s$ and $s'$ is the factor $\frac{s}{s+s'}$. For $f^{F(1)}_{qs}$, the only part which does not factorize in this way is the factor $\frac{1}{s-q}$.

For applications to black hole physics, one is interested in the limit of large $N_{1}N_{5}$. In this limit, component strings typically have parametrically large winding;
this is the main physical reason for studying the general $ M$, $ N $ problem.  Thus, as well as obtaining our exact results, we have extracted the behaviour of our results in the continuum limit, in which the mode numbers are large compared to the spacing of modes on the component string. We found significant simplification to the various quantities, similar to that observed in the bosonic case. These simplified expressions may prove useful in extracting the qualitative dynamics of thermalization in this theory. 

The results of this paper further our understanding of the effect of the deformation operator in the D1D5 CFT.  It is hoped that these results will lead to a better understanding of the process of thermalization in this theory, and thereby shed further light on aspects of black hole dynamics, in particular the processes of black hole formation and evaporation.

\b

\section*{Acknowledgements}

We thank Steve Avery, Borun Chowdhury, Amanda Peet and Ida Zadeh for discussions on various aspects of the deformation operator. This work was supported in part by DOE grant DE-FG02-91ER-40690. The work of D.T. was supported in part by the John Templeton Foundation Grant 48222: ``String Theory and the Anthropic Universe''.

\b

\begin{appendix}

\section{Notation and conventions} \label{ap:CFT-notation}

We follow the conventions of \cite{\acmtwo, \acmthree}, which we record here for convenience.
We have 4 real left moving fermions $\psi_1, \psi_2, \psi_3, \psi_4$ which we group into doublets $\psi^{\alpha A}$ as follows:
\be
\begin{pmatrix}
\psi^{++} \cr \psi^{-+}
\end{pmatrix}
=\sqi
\begin{pmatrix}
\psi_1+i\psi_2 \cr \psi_3+i\psi_4
\end{pmatrix}
\ee
\be
\begin{pmatrix}
\psi^{+-} \cr \psi^{--}
\end{pmatrix}
=\sqi
\begin{pmatrix}
\psi_3-i\psi_4 \cr -(\psi_1-i\psi_2)
\end{pmatrix}.
\ee
Here $\alpha=(+,-)$ is an index of the subgroup $SU(2)_L$ of rotations on $S^3$ and $A=(+,-)$ is an index of the subgroup $SU(2)_1$ from rotations in $T^4$. The reality conditions on the individual fermions are 
\bea
(\psi_i)^{\dagger} = \psi_i \qquad \Rightarrow \qquad (\psi^{\a A})^{\dagger} = - \epsilon_{\alpha\beta}\epsilon_{AB} \psi^{\beta B} \,.
\eea 
One can introduce doublets $\psi^\dagger$, whose components are given by 
\bea
 (\psi^\dagger)_{\alpha A} &=& (\psi^{\a A})^{\dagger} 
\eea 
in terms of which the reality condition is
\be
 (\psi^\dagger)_{\alpha A}=-\epsilon_{\alpha\beta}\epsilon_{AB} \psi^{\beta B} \,.
\ee
The 2-point functions are
\be
<\psi^{\alpha A}(z)(\psi^\dagger)_{\beta B}(w)>=\delta^\alpha_\beta\delta^A_B{1\over z-w}, ~~~
<\psi^{\alpha A}(z)\psi^{\beta B}(w)>=-\epsilon^{\alpha\beta}\epsilon^{AB}{1\over z-w}
\ee
where we have 
\be
\epsilon_{12}=1, ~~~\epsilon^{12}=-1, ~~~
\psi_A=\epsilon_{AB}\psi^B, ~~~
\psi^A=\epsilon^{AB}\psi_B \,.
\ee
There are 4 real left moving bosons $X_1, X_2, X_3, X_4$ which can be grouped into a matrix 
\be
X_{A\dot A}= \sqi X_i \sigma_i
=\sqi
\begin{pmatrix}
X_3+iX_4 & X_1-iX_2 \cr X_1+iX_2&-X_3+iX_4
\end{pmatrix}
\ee
where $\sigma_i=(\sigma_a, iI)$. The reality condition on the individual bosons is
\bea
(X_i)^{\dagger} = X_i \qquad \Rightarrow \qquad (X_{A\dot A})^{\dagger} = - \e^{AB}\e^{\dot A \dot B} X_{B \dot B} \,.
\eea 
One can introduce a matrix $X^\dagger$ with components  
\be
(X^\dagger)^{A\dot A}~=~ (X_{A\dot A})^{\dagger}~=~\sqi
\begin{pmatrix}
X_3-iX_4& X_1+iX_2\cr 
X_1-iX_2&-X_3-iX_4
\end{pmatrix}
\ee
in terms of which the reality condition is
\bea
(X^\dagger)^{A\dot A}~=~ - \e^{AB}\e^{\dot A \dot B} X_{B \dot B} \,.
\eea 
The 2-point functions are
\be
<\p X_{A\dot A}(z) (\p X^\dagger)^{B\dot B}(w)>=-{1\over (z-w)^2}\delta^B_A\delta^{\dot B}_{\dot A}, ~~~
<\p X_{A\dot A}(z) \p X_{B\dot B}(w)>={1\over (z-w)^2}\epsilon_{AB}\epsilon_{\dot A\dot B} \,.
\ee

The chiral algebra is generated by the operators
\be
J^a=-{1\over 4}(\psi^\dagger)_{\alpha A} (\sigma^{Ta})^\alpha{}_\beta \psi^{\beta A}
\ee
\be
G^\alpha_{\dot A}= \psi^{\alpha A} \p X_{A\dot A}, ~~~(G^\dagger)_{\alpha}^{\dot A}=(\psi^\dagger)_{\alpha A} \p (X^\dagger)^{A\dot A}
\ee
\be
T=-{1\over 2} (\p X^\dagger)^{A\dot A}\p X_{A\dot A}-{1\over 2} (\psi^\dagger)_{\alpha A} \p \psi^{\alpha A}
\ee
\be
(G^\dagger)_{\alpha}^{\dot A}=-\epsilon_{\alpha\beta} \epsilon^{\dot A\dot B}G^\beta_{\dot B}, ~~~~G^{\alpha}_{\dot A}=-\epsilon^{\alpha\beta} \epsilon_{\dot A\dot B}(G^\dagger)_\beta^{\dot B} \,.
\ee
These operators generate the algebra
\be
J^a(z) J^b(z')\sim \delta^{ab} {\h\over (z-z')^2}+i\epsilon^{abc} {J^c\over z-z'}
\ee
\be
J^a(z) G^\alpha_{\dot A} (z')\sim {1\over (z-z')}\h (\sigma^{aT})^\alpha{}_\beta G^\beta_{\dot A}
\ee
\be
G^\alpha_{\dot A}(z) (G^\dagger)^{\dot B}_\beta(z')\sim -{2\over (z-z')^3}\delta^\alpha_\beta \delta^{\dot B}_{\dot A}- \delta^{\dot B}_{\dot A}  (\sigma^{Ta})^\alpha{}_\beta [{2J^a\over (z-z')^2}+{\p J^a\over (z-z')}]
-{1\over (z-z')}\delta^\alpha_\beta \delta^{\dot B}_{\dot A}T
\ee
\be
T(z)T(z')\sim {3\over (z-z')^4}+{2T\over (z-z')^2}+{\p T\over (z-z')}
\ee
\be
T(z) J^a(z')\sim {J^a\over (z-z')^2}+{\p J^a\over (z-z')} 
\ee
\be
T(z) G^\alpha_{\dot A}(z')\sim {{3\over 2}G^\alpha_{\dot A}\over (z-z')^2}  + {\p G^\alpha_{\dot A}\over (z-z')} \,.
\ee

Note that
\be
J^a(z) \psi^{\gamma C}(z')\sim {1\over 2} {1\over z-z'} (\sigma^{aT})^\gamma{}_\beta \psi^{\beta C} \,.
\ee

The above OPE algebra gives the commutation relations
\begin{eqnarray}
\com{J^a_m}{J^b_n} &=& \frac{m}{2}\delta^{ab}\delta_{m+n,0} + i{\epsilon^{ab}}_c J^c_{m+n}
            \\
\com{J^a_m}{G^\alpha_{\dot{A},n}} &=& \frac{1}{2}{(\sigma^{aT})^\alpha}_\beta G^\beta_{\dot{A},m+n}
             \\
\ac{G^\alpha_{\dot{A},m}}{G^\beta_{\dot{B},n}} &=& \hspace*{-4pt}\epsilon_{\dot{A}\dot{B}}\bigg[
   (m^2 - \frac{1}{4})\epsilon^{\alpha\beta}\delta_{m+n,0}
  + (m-n){(\sigma^{aT})^\alpha}_\gamma\epsilon^{\gamma\beta}J^a_{m+n}
  + \epsilon^{\alpha\beta} L_{m+n}\bigg]\quad\\
\com{L_m}{L_n} &=& \frac{m(m^2-\frac{1}{4})}{2}\delta_{m+n,0} + (m-n)L_{m+n}\\
\com{L_m}{J^a_n} &=& -n J^a_{m+n}\\
\com{L_m}{G^\alpha_{\dot{A},n}} &=& \left(\frac{m}{2}-n\right)G^\alpha_{\dot{A},m+n} \,.
\end{eqnarray}

\section{Mapping modes to the covering plane}\label{sec:mapping}

In this appendix we compute the transformations of the fermion modes  under the sequence of spectral flow transformations and coordinate changes described in Section \ref{sec:sf}. 

\subsection{Modes on the cylinder}

We start by recalling from Section \ref{sec:expansions} the mode expansions on the cylinder. 

Below the twist, we have:
\begin{eqnarray}
d_m^{(1)\alpha A} &=& \frac{1}{2\pi i\sqrt M} \int\limits_{\sigma=0}^{2\pi M} \psi^{(1)\alpha A} (w) e^{\frac{m}{M}w} dw\\
d_m^{(2)\alpha A} &=&  \frac{1}{2\pi i\sqrt N} \int\limits_{\sigma=0}^{2\pi N} \psi^{(2)\alpha A} (w) e^{\frac{m}{N}w} dw,
\end{eqnarray}
while above the twist, we have:
\begin{eqnarray}
d_k^{\alpha A} &=& \frac{1}{2\pi i\sqrt{M+N}} \int\limits_{\sigma=0}^{2\pi (M+N)} \psi^{\alpha A} (w) e^{\frac{k}{M+N}w} dw,
\end{eqnarray}
which gives
\begin{eqnarray}
\psi^{\alpha A}\left(w\right)=\frac{1}{\sqrt{M+N}}\sum_{k}d_{k}^{\alpha A}e^{-\frac{k}{M+N}w}.
\end{eqnarray}
The anti-commutation relations are
\begin{eqnarray}
\left\{d_k^{\alpha A}, d_l^{\beta B}\right\} &=& -\varepsilon^{\alpha\beta} \varepsilon^{AB} \delta_{k+l,0}.
\end{eqnarray}

\subsection{Modes on the $z$ plane}

We map the cylinder to the plane with coordinate $z$ via
\begin{eqnarray}
z &=& e^w.
\end{eqnarray}
The operator modes transform as follows. Before the twist, i.e. $|z| < e^{\tau_0}$, using a contour circling $z=0$, we have
\begin{eqnarray}
d_m^{(1)\alpha A} & \to & \frac{1}{2\pi i \sqrt{M}} \int\limits^{2\pi M}_{\arg (z) = 0} \psi ^{(1) \alpha A} (z) z^{\frac{m}{M} - \frac12} dz\cr
d_m^{(2)\alpha A} & \to & \frac{1}{2\pi i\sqrt{N}} \int\limits_{\arg (z) = 0}^{2\pi N} \psi^{(2)\alpha A} (z) z^{\frac{m}{N}-\frac12} dz.
\end{eqnarray}
After the twist, i.e. $|z| > e^{\tau_0}$, using a contour circling $z=\infty$, we have
\begin{eqnarray}
d_k^{\alpha A} &\to& \frac{1}{2\pi i \sqrt{M + N}} \int\limits_{\arg (z) = 0}^{2\pi (M+N)} \psi ^{\alpha A}(z) z^{\frac{k}{M+N} - \frac{1}{2}} dz.
\end{eqnarray}

\subsection{Modes on the covering space}

We now map the problem to the covering space with coordinate $t$, where the fields are single-valued.  We use the map
\begin{eqnarray}
z &=& t^M (t-a)^N.
\end{eqnarray}
It is convenient to write the derivative as
\begin{eqnarray}
\frac{dz}{dt} &=& (M+N) \frac{z}{t(t-a)} \left(t -\tfrac{M}{M+N}a\right).
\end{eqnarray}
Under this map, the modes before the twist become:
\begin{eqnarray}
d_m^{(1)\alpha A} &\to & \frac{\sqrt{M+N}}{2\pi i\sqrt{M}} \oint\limits_{t = 0} dt \, \psi^{\alpha A} (t) \left[t^{m-\frac12} \left(t - \tfrac{M}{M+N}a\right)^{\frac12} (t-a)^{\frac{Nm}{M} - \frac12}\right]\cr
d_m^{(2)\alpha A} &\to & \frac{\sqrt{M+N}}{2\pi i\sqrt{N}} \oint\limits_{t = a} dt \, \psi^{\alpha A} (t) \left[t^{\frac{Mm}{N}-\frac12} \left(t - \tfrac{M}{M+N}a\right)^{\frac12} (t-a)^{m-\frac12}\right]
\end{eqnarray}
and after the twist, we have:
\begin{eqnarray}
d_k^{\alpha A} &\to&  \frac{1}{2\pi i} \oint\limits_{t = \infty} dt \, \psi^{\alpha A} (t) \left[t^{\frac{Mk}{M+N}-\frac12} \left(t - \tfrac{M}{M+N}a\right)^{\frac12} (t-a)^{\frac{Nk}{M+N} - \frac12}\right].
\end{eqnarray}

\subsection{Modes after the first spectral flow}

We now perform the sequence of spectral flow transformations and coordinate changes described in Section \ref{sec:sf}.
As explained there, at this point we ignore normalization factors of spin fields and transformation properties of spin fields under spectral flow. We spectral flow in the same way on left- and right-moving sectors,  but write only the holomorphic expressions. 

We first spectral flow by $\alpha = 1$ in the $t$ plane. Before the spectral flow, we have the amplitude
\begin{eqnarray}
&&{}_t \bra{0_{R,-}} S^- (a) S^+ \left(\tfrac{M}{M+N}a\right)\ket{0_R^-}_t.
\end{eqnarray}
The spectral flow has the following effects:
\begin{enumerate}
\item[(a)] The R ground states $\ket{0_R^-}_t$ and ${}_t\bra{0_{R,-}}$ map to the NS ground states in the $t$ plane:
\be
\ket{0_R^-}_t \to \ket{0_{NS}}_t, \qquad {}_t\bra{0_{R,-}} \to {}_t\bra{0_{NS}}
\ee
\item[(b)] The modes of the fermion fields change as follows (the bosonic modes are unaffected):
    \begin{eqnarray}
    \psi^{\pm A} (t)&\to &t^{\mp\frac12} \psi^{\pm A}(t).
    \end{eqnarray}
\end{enumerate}
Then the modes before the twist become:
    \begin{eqnarray}
    d_m^{(1) + A} &\to& \frac{\sqrt{M+N}}{2\pi i \sqrt M} \oint\limits_{t = 0} {\rm dt} \psi^{(1)+A} (t) \left[t^{m-1} \left(t - \tfrac{M}{M+N}a\right) ^{\frac12} (t-a)^{\frac{Nm}{M} - \frac12}\right]\cr
        d_m^{(2) + A} &\to& \frac{\sqrt{M+N}}{2\pi i \sqrt N} \oint\limits_{t = a} {\rm dt} \psi^{(2)+A} (t) \left[t^{\frac{Mm}{N}-1} \left(t - \tfrac{M}{M+N}a\right) ^{\frac12} (t-a)^{m - \frac12}\right]\cr
    d_m^{(1) - A} &\to& \frac{\sqrt{M+N}}{2\pi i \sqrt M} \oint\limits_{t = 0} {\rm dt} \psi^{(1)-A} (t) \left[t^{m} \left(t - \tfrac{M}{M+N}a\right) ^{\frac12} (t-a)^{\frac{Nm}{M} - \frac12}\right]        \cr
    d_m^{(2) - A} &\to& \frac{\sqrt{M+N}}{2\pi i \sqrt N} \oint\limits_{t = a} {\rm dt} \psi^{(2)-A} (t) \left[t^{\frac{Mm}{N}} \left(t - \tfrac{M}{M+N}a\right) ^{\frac12} (t-a)^{m - \frac12}\right]
    \end{eqnarray}
and the modes after the twist become:
    \begin{eqnarray}
        d_k^{ + A} &\to& \frac{1}{2\pi i } \oint\limits_{t = \infty} {\rm dt} \psi^{+A} (t) \left[t^{\frac{Mk}{M+N}-1} \left(t - \tfrac{M}{M+N}a\right) ^{\frac12} (t-a)^{\frac{Nk}{M+N} - \frac12}\right]\cr
         d_k^{ - A} &\to & \frac{1}{2\pi i} \oint\limits_{t = \infty} {\rm dt} \psi^{-A} (t) \left[t^{\frac{Mk}{M+N}} \left(t - \tfrac{M}{M+N}a\right) ^{\frac12} (t-a)^{\frac{Nk}{M+N} - \frac12}\right].
\end{eqnarray}

\subsection{Modes after the second spectral flow}
Next, we change coordinate to
\begin{eqnarray}
t'& =& t - \tfrac{M}{M+N}a,
\end{eqnarray}
and we spectral flow by $\alpha = -1$ in the $t'$ plane. Before this second spectral flow, we have the amplitude
\begin{eqnarray}
&&_{t'} \bra{0_{NS}} S^- \left(\tfrac{N}{M+N} a\right) \ket{0_R^+} _{t'}.
\end{eqnarray}
The spectral flow has the following effects:
\begin{enumerate}
\item[(a)] The R ground state $\ket{0_R^+}_{t'}$ maps to the NS vacuum in the $t'$ plane:
\begin{eqnarray}
\ket{0_R^+}_{t'} &\to &\ket{0_{NS}}_{t'}.
\end{eqnarray}
\item[(b)] The modes of the fermions change as follows:
\begin{eqnarray}
\psi^{\pm A} (t') &\to& (t')^{\pm \frac12} \psi^{\pm A} (t').
\end{eqnarray}
\end{enumerate}
Then the modes before the twist become:
  \begin{eqnarray}
  d_{m}^{\left(1\right)+ A} &\to& 	\frac{\sqrt{M+N}}{2\pi i \sqrt{M}}
	\oint\limits_{t'=-\frac{Ma}{M+N}} dt' \psi^{\left(1\right)+A}
	\left[\left(t'+\tfrac{M}{M+N}a\right)^{m-1}t'\left(t'-\tfrac{N}{M+N}a\right)^{\frac{Nm}{M}-\frac{1}{2}}\right]\cr
	d_{m}^{\left(2\right)+ A} &\to& \frac{\sqrt{M+N}}{2\pi i \sqrt{N}}
	\oint\limits_{t'=\frac{Na}{M+N}} dt' \psi^{\left(2\right)+A}
	\left[\left(t'+\tfrac{M}{M+N}a\right)^{\frac{Mm}{N}-1}t'\left(t'-\tfrac{N}{M+N}a\right)^{m-\frac{1}{2}}\right]\cr
 	d_{m}^{\left(1\right)-A}&\to&\frac{\sqrt{M+N}}{2\pi i \sqrt{M}}
	\oint\limits_{t'=-\frac{Ma}{M+N}} dt'\psi^{\left(1\right)-A}
	\left[\left(t'+\tfrac{M}{M+N}a\right)^{m}\left(t'-\tfrac{N}{M+N}a\right)^{\frac{Nm}{M}-\frac{1}{2}}\right]\cr
 	d_{m}^{\left(2\right)-A}&\to&\frac{\sqrt{M+N}}{2\pi i \sqrt{N}}
	\oint\limits_{t'=\frac{Na}{M+N}} dt'\psi^{\left(2\right)-A}
	\left[\left(t'+\tfrac{M}{M+N}a\right)^{\frac{Mm}{N}}\left(t'-\tfrac{N}{M+N}a\right)^{m-\frac{1}{2}}\right]
  \end{eqnarray}
and the modes after the twist become:
\begin{eqnarray}
       d_{k}^{+A}&\to& \frac{1}{2 \pi i}\oint\limits_{t'=\infty}\psi^{+A}\left(t'\right)\left[\left(t'+\tfrac{M}{M+N}a\right)^{\frac{Mk}{M+N}-1}t'\left(t'-\tfrac{N}{M+N}a\right)^{\frac{Nk}{M+N}-\frac12}\right]\cr
      d_{k}^{-A}&\to& \frac{1}{2 \pi i}\oint\limits_{t'=\infty}\psi^{-A}\left(t'\right)\left[\left(t'+\tfrac{M}{M+N}a\right)^{\frac{Mk}{M+N}}\left(t'-\tfrac{N}{M+N}a\right)^{\frac{Nk}{M+N}-\frac12}\right].
\end{eqnarray}

\subsection{Modes after the third spectral flow}
Next, we change the coordinate to
\begin{eqnarray}
&&\hat{t}=t'-\tfrac{N}{M+N}a,
\end{eqnarray}
and we spectral flow by $\alpha =1$ in the $\hat t$ plane. The spectral flow has the following effects
\begin{enumerate}
\item[(a)] The R ground state maps to the NS vacuum in the $\hat{t}$ plane:
\begin{eqnarray}
\ket{0_{R}^{-}}_{\hat{t}}&\to& \ket{0_{NS}}_{\hat{t}}.
\end{eqnarray}
\item[(b)] The modes of the fermions change as follows:
\begin{eqnarray}
\psi^{\pm A}\left(\hat{t}\;\!\right)&\to&\left(\hat{t}\;\!\right)^{\mp\frac12}\psi^{\pm A}\left(\hat{t}\;\!\right)
\end{eqnarray}
\end{enumerate}
Then the modes before the twist become:
\begin{eqnarray}
    d_{m}^{\left(1\right)+A}&\to& \hat{d}_{m}'^{\left(1\right)+A}
		~=~\frac{\sqrt{M+N}}{2\pi i\sqrt{M}}
		\oint\limits_{\hat{t}=-a}d\hat{t}\,\psi^{\left(1\right)+A}\left(\hat{t}\;\!\right)\left[\left(\hat{t}+a\right)^{m-1}\left(\hat{t}+\tfrac{Na}{M+N}\right)\hat{t}^{\frac{Nm}{M}-1}\right]\cr
     d_{m}^{\left(2\right)+A}&\to& \hat{d}_{m}'^{\left(2\right)+A}
		~=~\frac{\sqrt{M+N}}{2\pi i\sqrt{N}}
		\oint\limits_{\hat{t}=0}d\hat{t}\,\psi^{\left(2\right)+A}\left(\hat{t}\;\!\right)\left[\left(\hat{t}+a\right)^{\frac{Mm}{N}-1}\left(\hat{t}+\tfrac{Na}{M+N}\right)\hat{t}^{m-1}\right]\cr
     d_{m}^{\left(1\right)-A}&\to& \hat{d}_{m}'^{\left(1\right)-A}
		~=~\frac{\sqrt{M+N}}{2\pi i\sqrt{M}}
		\oint\limits_{\hat{t}=-a}d\hat{t}\,\psi^{\left(1\right)-A}\left(\hat{t}\;\!\right)\left[\left(\hat{t}+a\right)^{m}\hat{t}^{\frac{Nm}{M}}\right]\cr
     d_{m}^{\left(2\right)-A}&\to& \hat{d}_{m}'^{\left(2\right)-A}
		~=~\frac{\sqrt{M+N}}{2\pi i\sqrt{N}}
		\oint\limits_{\hat{t}=0}d\hat{t}\,\psi^{\left(2\right)-A}\left(\hat{t}\;\!\right)\left[\left(\hat{t}+a\right)^{\frac{Mm}{N}}\hat{t}^{m}\right]
\end{eqnarray}
After the twist, we obtain
\begin{eqnarray}
   d_{k}^{+A}&\to&\hat{d}_{k}'^{+A}~=~\frac{1}{2\pi i}\oint\limits_{\hat{t}=\infty}d\hat{t}\,\psi^{+A}\left(\hat{t}\;\!\right)\left[\left(\hat{t}+a\right)^{\tfrac{Mk}{M+N}-1}\left(\hat{t}+\tfrac{N}{M+N}a\right)\hat{t}^{\frac{Nk}{M+N}-1}\right]\cr
   d_{k}^{-A}&\to&\hat{d}_{k}'^{-A}~=~\frac{1}{2\pi i}\oint\limits_{\hat{t}=\infty}d\hat{t}\,\psi^{-A}\left(\hat{t}\;\!\right)\left[\left(\hat{t}+a\right)^{\tfrac{Mk}{M+N}}\hat{t}^{\frac{Nk}{M+N}}\right].
\label{dhat-a}
\end{eqnarray}

\b

\section{Calculation of the overall prefactor $C_{MN}$} \label{app:cmn}

In this appendix we compute the prefactor $C_{MN}$ given by
\bea
C_{MN}&=&\bra{0_{R,--}}
\sigma_{2}^{++}\left(w_{0}\right)
\ket{0_{R}^{--}}^{\left(1\right)}\ket{0_{R}^{--}}^{\left(2\right)} \,.
\label{eq:cmn-a}
\eea
We will have four contributions to our factor of $C_{MN}$. 

First we lift the correlator to the $ z $ plane. The conformal weight of the twist operator then gives a Jacobian factor, which is our first contribution. 

We then have the $ z $ plane correlator
\begin{eqnarray}
\bra{0_{R,--}}
\sigma_{2}^{++}\left(z_{0}\right)
\ket{0_{R}^{--}}^{\left(1\right)}\ket{0_{R}^{--}}^{\left(2\right)} 
&=&
\langle \sigma_{M+N}^{++}(\infty)\sigma_{2}^{++}(z_{0})\sigma_{N}^{--}(0)\sigma_{M}^{--}(0)\rangle\,.
\end{eqnarray}
We employ the methods developed in \cite{\lmone,\lmtwo} to compute this correlator.
Introducing the notation that $\sigma_{M+N}^{++}$ has dimension $(\Delta_{{M+N}}^{++},\Delta_{{M+N}}^{++})$, we have
\begin{eqnarray}
\langle\sigma_{M+N}^{++}(\infty)\sigma_{2}^{++}(z_{0})\sigma_{N}^{--}(0)\sigma_{M}^{--}(0)\rangle
&\equiv&
\lim_{|z|\to\infty}|z|^{4\Delta_{{M+N}}^{++}}
\langle\sigma_{M+N}^{++}(z)\sigma_{2}^{++}(z_{0})\sigma_{N}^{--}(0)\sigma_{M}^{--}(0)\rangle
\cr
&=& \lim_{|z|\to\infty}
\frac{
\langle\sigma_{M+N}^{++}(z)\sigma_{2}^{++}(z_{0})\sigma_{N}^{--}(0)\sigma_{M}^{--}(0)\rangle
}
{
\langle\sigma_{M+N}^{++}(z)\sigma_{M+N}^{--}(0)\rangle}\,.
\eea
The computation of this ratio of correlators factorizes~\cite{\lmtwo} into a product of terms coming from the Liouville action, which depend only on the `bare twist' part of the above spin-twist fields, and 
$t$ plane correlators of appropriately normalized spin fields. We denote the image of $z$ and $z_0$ in the $t$ plane by $t(z)$ and $t_0$ respectively. Then we have
\begin{eqnarray}
\langle \sigma_{M+N}^{++}(\infty)\sigma_{2}^{++}(z_{0})\sigma_{N}^{--}(0)\sigma_{M}^{--}(0)\rangle
&=& \lim_{|z|\to\infty}
\frac{\langle\sigma_{M+N}(z)\sigma_{2}(z_{0})\sigma_{N}(0)\sigma_{M}(0)\rangle}
{\langle\sigma_{M+N}(z)\sigma_{M+N}(0)\rangle}
\frac{\langle \mathcal{S}_4 (t(z),t_0) \rangle}{\langle \mathcal{S}_2 (t(z))\rangle} \cr &&
\label{eq:main}
\end{eqnarray}
where $\langle \mathcal{S}_4 (t(z),t_0) \rangle$ and $\langle \mathcal{S}_2 (t(z))\rangle$ are  $t$ plane correlators (four-point and two-point respectively) of appropriately normalized spin fields.

Our second contribution will be the ratio of correlators of bare twist fields in \eq{eq:main}; our third contribution will be the various normalization factors for the $t$ plane spin field correlators, and our final contribution will be the ratio of $t$ plane spin field correlators.

We now compute these four contributions  in turn.

\subsection{Jacobian factor for lifting from cylinder to plane}

We first lift the correlator to the $z$ plane. Since $\sigma_{2}^{++}$ has weight (1/2,1/2), we obtain the Jacobian factor contribution
\begin{eqnarray}
\left|\frac{dz}{dw}\right|_{z=z_{0}}&=&|a|^{M+N}\frac{M^{M}N^{N}}{(M+N)^{M+N}} \,.
\label{Jacobian}
\end{eqnarray}

\subsection{Liouville action terms}

In this subsection we deal only with the contribution from the bare twist fields, i.e. we compute
\bea
\langle\sigma_{M+N}(\infty)\sigma_{2}(z_{0})\sigma_{N}(0)\sigma_{M}(0)\rangle
&=&
\lim_{|z|\to\infty}
\frac{\langle\sigma_{M+N}(z)\sigma_{2}(z_{0})\sigma_{N}(0)\sigma_{M}(0)\rangle}
{\langle\sigma_{M+N}(z)\sigma_{M+N}(0)\rangle}
\label{eq:cmn-a3}
\eea
where the bare twist operators are normalized as
\begin{eqnarray}
\langle\sigma_{M}(z)\sigma_{M}(0)\rangle=\frac{1}{|z|^{4\Delta_{M}}} \,.
\label{norm}
\end{eqnarray}

\subsubsection{Defining regulated twist operators}

We work in a path integral formulation, and we
define regularized bare twist operators by cutting a small circular hole of radius $\epsilon \ll 1$ around each finite insertion point,
\begin{eqnarray} 
\sigma^{\epsilon}_{2}(z_{0}) \,, ~~\sigma^{\epsilon}_{M}(0) \,,~~\sigma^{\epsilon}_{N}(0) \,.
\end{eqnarray}
From \eq{norm} we have the following relationship between the regularized twist operators (whose normalization depends on the cutoff) and the canonically normalized twist operators:
\begin{eqnarray}
\sigma_{M}=\frac{1}{\sqrt{\langle\sigma_{M}^{\epsilon}(0)\sigma_{M}^{\epsilon}(1)\rangle}}\sigma_{M}^{\epsilon} \,.
\label{twist_norm}
\end{eqnarray}
We also define a regularized twist operator at infinity by cutting a large circular hole of radius $1/\t\delta$ with $\t\delta \ll 1$. This operator is denoted
\begin{eqnarray} 
\sigma^{\t\delta}_{M+N}(\infty) \,.
\end{eqnarray}
Since in \eq{eq:cmn-a3} we have a ratio of amplitudes, we do not need to worry about the normalization of $\sigma^{\t\delta}_{M+N}(\infty) $, and we obtain
\begin{eqnarray}
\lim_{|z|\to\infty}
\frac{\langle\sigma_{M+N}(z)\sigma_{2}(z_{0})\sigma_{N}(0)\sigma_{M}(0)\rangle}
{\langle\sigma_{M+N}(z)\sigma_{M+N}(0)\rangle}
&=&\frac{\langle\sigma^{\tilde{\delta}}_{M+N}(\infty)\sigma_{2}(z_{0})\sigma_{N}(0)\sigma_{M}(0)\rangle}{\langle\sigma_{M+N}^{\tilde{\delta}}(\infty)\sigma_{M+N}(0)\rangle} \,.
\end{eqnarray}
From (\ref{twist_norm}) we then obtain
\begin{eqnarray}
&&\langle\sigma_{M+N}(\infty)\sigma_{2}(z_{0})\sigma_{N}(0)\sigma_{M}(0)\rangle=\cr
&&\quad\frac{\langle\sigma^{\tilde{\delta}}_{M+N}(\infty)\sigma_{2}^{\epsilon}(z_{0})\sigma_{N}^{\epsilon}(0)\sigma_{M}^{\epsilon}(0)\rangle}{\langle\sigma_{M+N}^{\tilde{\delta}}(\infty)\sigma_{M+N}^{\epsilon}(0)\rangle}
\sqrt{
\frac{\langle\sigma_{M+N}^{\epsilon}(0)\sigma_{M+N}^{\epsilon}(1)\rangle}
{\langle\sigma_{2}^{\epsilon}(0)\sigma_{2}^{\epsilon}(1)\rangle
\langle\sigma_{N}^{\epsilon}(0)\sigma_{N}^{\epsilon}(1)\rangle
\langle\sigma_{M}^{\epsilon}(0)\sigma_{M}^{\epsilon}(1)\rangle}
} \, \qquad
\label{eq:4pt-1}
\end{eqnarray}
and so our four-point function is now expressed entirely in terms of correlators of regularized twist operators, which we now compute.

\subsubsection{Lifting to the $t$ plane}

We now lift the problem to the $t$-plane using the map \eq{eq:covermap},
\bea\label{eq:covermap-a}
z &=& t^M(t-a)^N \,.
\eea
In the $ z $ plane we have cut out various circular holes, leaving a path integral over an open set. The image of this open set in the $ t $ plane will be denoted by $\Sigma$. In the $ t $ plane,  we have single-valued fields, with no twist operators, but with appropriately normalized spin field insertions (which we deal with later). There is also a non-trivial metric, which we deal with in the present section.

To take account of the non-trivial metric on the $ t $ plane, we define a fiducial metric on the cover space $\Sigma$ and compute the Liouville action. Replacing the $z$-plane by a closed surface (a sphere), we define the $z$-plane metric $g$ via
\begin{eqnarray}
ds^{2}&=&\begin{cases}
dzd\bar{z} & |z|<\frac{1}{\delta}\\
d\tilde{z}d\bar{\tilde{z}} & |\tilde{z}|<\frac{1}{\delta}
\end{cases} \nn
\tilde{z}&=&\frac{1}{\delta^{2}}\frac{1}{z}
\label{z_fid}
\end{eqnarray}
where $\delta$ is a cutoff which will not play a role in our computation.

We define the fiducial metric on the $ t $ plane $\hat{g}$, to be flat so the curvature term in the Liouville action vanishes. We define $\hat{g}$ via 
\begin{eqnarray}
d\hat{s}^{2}&=& \begin{cases}
dtd\bar{t} & |t|< \frac{1}{\delta'}\\
d\tilde{t}d\bar{\tilde{t}} & |\tilde{t}|< \frac{1}{\delta'}
\end{cases}\nn
\tilde{t}&=&\frac{1}{\delta'^{2}}\frac{1}{t} \,.
\end{eqnarray} 
where $\delta'$ is another cutoff which will not play a role in our computation.

In lifting to the $t$-plane from the $z$-plane our metric will change by
\begin{eqnarray}
ds^{2}=e^{\phi}d\hat{s}^{2}
\label{metric_anomaly}
\end{eqnarray}
This corresponds to a transformation of our path integral of the form
\begin{eqnarray}
Z^{(g)}=e^{S_{L}}Z^{(\hat{g})}
\label{partition_trans}
\end{eqnarray}
where $S_{L}$ is the Liouville action given by (see e.g.~\cite{Friedan:1982is})
\begin{eqnarray}
S_{L}=\frac{c}{96\pi}\int d^{2}t\sqrt{-\hat g}\left[\partial_{\mu}\phi\partial_{\nu}\phi \,{\hat g}^{\m\n}+2R^{(\hat{g})}\phi\right].
\end{eqnarray}
Since we map from the $z$ space to the $t$ space we have an induced metric given by
\begin{eqnarray}
ds^{2}=dzd\bar{z}=\frac{dz}{dt}\frac{d\bar{z}}{d\bar{t}}dtd\bar{t}
\end{eqnarray}
Comparing this equation with (\ref{metric_anomaly}) we find that the Liouville field is given by
\begin{eqnarray}
\phi=\log\frac{dz}{dt}+\log\frac{d\bar{z}}{d\bar{t}} = \log \left ( \left | {dz\over dt} \right |^2 \right ).
\end{eqnarray}

Since the fiducial metric is flat and $\partial_{\mu}\partial^{\mu}\phi=0$, the Liouville action becomes a boundary term, which can be written as
\begin{eqnarray}
S_{L}=\frac{c}{96\pi}\left[i\int\limits_{\partial} dt\phi\partial_{t}\phi+c.c\right]
\label{eq:sl}
\end{eqnarray}
where $c.c.$ denotes the complex conjugate and where $\partial$ is the boundary of $\Sigma$, comprising the images in the $t$ plane of the circular holes defined in the $z$ plane.

The contribution of correlation function reduces to calculating the contributions to the Liouville action from the various holes comprising the boundary of $\Sigma$, which in our problem are:
\begin{itemize}
	\item \underline{Type 1}: The holes whose images are at finite points in the $t$-plane.  In our problem these holes are from finite points within the $z$-plane. 
	\item \underline{Type 2}: The hole whose image is at $t=\infty$.  This hole comes from a cut in the $z$-plane at $\frac{1}{\tilde{\delta}}$ upon taking $\tilde{\delta}\to 0$.  Since all $M+N$ copies of the CFT are twisted together at $z=\infty$, this is the only hole that appears at $t = \infty$.
\end{itemize}

Then our four-point correlation function \eq{eq:4pt-1} can be written as
\begin{eqnarray}
&&\langle\sigma_{M+N}(\infty)\sigma_{2}(z_{0})\sigma_{N}(0)\sigma_{M}(0)\rangle
=\cr
&&\qquad\frac{e^{S_{L_{1}}[\sigma_{M+N}^{\tilde{\delta}}\sigma_{2}^{\epsilon}\sigma_{N}^{\epsilon}\sigma_{M}^{\epsilon}]
+S_{L_{2}}[\sigma_{M+N}^{\tilde{\delta}}\sigma_{2}^{\epsilon}\sigma_{N}^{\epsilon}\sigma_{M}^{\epsilon}]}}
{e^{S_{L_{1}}[\sigma_{M+N}^{\tilde{\delta}}\sigma_{M+N}^{\epsilon}]
+S_{L_{2}}[\sigma_{M+N}^{\tilde{\delta}}\sigma_{M+N}^{\epsilon}]}}
\sqrt{
\frac{\langle\sigma_{M+N}^{\epsilon}(0)\sigma_{M+N}^{\epsilon}(1)\rangle}
{\langle\sigma_{2}^{\epsilon}(0)\sigma_{2}^{\epsilon}(1)\rangle
\langle\sigma_{N}^{\epsilon}(0)\sigma_{N}^{\epsilon}(1)\rangle
\langle\sigma_{M}^{\epsilon}(0)\sigma_{M}^{\epsilon}(1)\rangle}
}\qquad\quad
\label{eq:4pt-2}
\end{eqnarray}
where $S_{L_{1}}[\sigma_{M+N}^{\tilde{\delta}}\sigma_{2}^{\epsilon}\sigma_{N}^{\epsilon}\sigma_{M}^{\epsilon}]$ and $S_{L_{2}}[\sigma^{\tilde{\delta}}_{M+N}\sigma_{2}^{\epsilon}\sigma_{N}^{\epsilon}\sigma_{N}^{\epsilon}]$ are the Liouville contributions to the four-point function from holes of Type 1 and Type 2 respectively, and similarly for the two-point function.

\subsubsection{The four-point function of regularized twist operators}

We now calculate the contribution to the Liouville action from the four-point function of regularized twist operators,
\begin{equation}
S_{L,4} ~\equiv~S_{L_{1}}[\sigma_{M+N}^{\tilde{\delta}}\sigma_{2}^{\epsilon}\sigma_{N}^{\epsilon}\sigma_{M}^{\epsilon}]
+S_{L_{2}}[\sigma_{M+N}^{\tilde{\delta}}\sigma_{2}^{\epsilon}\sigma_{N}^{\epsilon}\sigma_{M}^{\epsilon}]\,.
\end{equation}

Let us begin by looking at the contribution to the Liouville action from the branch point at $z=0$, $t=0$.  Near this point we have:
\begin{equation}
z=t^{M}(t-a)^{N}\approx t^{M}(-a)^{N};\,\,\,\,\,\; \frac{dz}{dt}\approx Mt^{M-1}(-a)^{N};\,\,\,\,\,\,t\approx \left (\frac{z}{(-a)^{N}}\right )^{1/M};
\end{equation}
\begin{equation}
\phi = \log\left [ \frac{dz}{dt}\right ] +c.c. \approx 2\log \left [ M|t|^{M-1}|a|^N \right ]; \,\,\,\,\,\, \partial_{t}\phi \approx \frac{M-1}{t}
\end{equation}
From \eq{eq:sl}, the contribution to the Liouville action from this point is given by (we note that the central charge is $c=6$)
\begin{equation}
S_{L_1}(z=0,t=0)=\frac{6}{96\pi}\left [ i\int dt \phi\partial_{t}\phi + c.c. \right ]
\end{equation}
We perform branch point integration introducing the following coordinates
\begin{equation}
z \approx \epsilon e^{i\theta}; \qquad t\approx \left (\frac{\epsilon}{(-a)^N}\right )^{1/M}e^{i\theta'}; \qquad \theta' = {\theta \over M}
\end{equation}
Making this change of variables for the integral, we obtain the four-point function's contribution to the Liouville action from this region:
\begin{equation}
\begin{split}
S_{L_1,4}(z=0,t=0)&=\frac{6}{96\pi}\left [ i\int\limits_{0}^{2\pi}d\theta' it\,\frac{M-1}{t}2\log[M|t|^{M-1}|a|^N]+c.c. \right ]\\
&=-\frac{1}{2}(M-1)\left (\log \left [ M\epsilon^{\frac{M-1}{M}}|a|^{\frac{N}{M}}\right ] \right )
\end{split}
\end{equation}

The contributions from the other branch points in the four-point function are calculated in the same manner.  Here we present only the results.
\begin{eqnarray}
S_{L_1,4}(z=0,t=a)&=&-\frac{1}{2}(N-1)\left ( \log \left [ N\epsilon^{\frac{N-1}{N}}|a|^{\frac{M}{N}}\right ] \right )\nn
S_{L_1,4}(z=z_{0},t=t_{0})&=&-\frac{1}{4}\log\left [ 4\left | b_{t_0}\right | \epsilon \right ] \nn
S_{L_2,4}(z=\infty,t=\infty)&=&\frac{1}{2}(M+N-1)\log\left [ (M+N)\tilde{\delta}^{-\frac{M+N-1}{M+N}} \right ]
\end{eqnarray}
where 
\begin{equation}\label{bt0}
\left | b_{t_0}\right |=\frac{|a|^{M+N-2}}{2}\frac{M^{M}N^{N}}{(M+N)^{M+N}}\left ( \frac{(M+N)^{3}}{MN}\right ).
\end{equation}

Combining the Liouville terms for the regularized four-point function we find a total contribution of
\begin{eqnarray} \label{Liouville4}
S_{L,4}&=&-\frac{1}{2}(M-1)\log \left [ M\epsilon^{\frac{M-1}{M}}|a|^{\frac{N}{M}} \right ]-\frac{1}{2}(N-1)\log \left [ N\epsilon^{\frac{N-1}{N}}|a|^{\frac{M}{N}} \right ]-\frac{1}{4}\log\left [ 4\left | b_{t_0}\right | \epsilon\right ]\nn
&&\quad{}+\frac{1}{2}(M+N-1)\log \left [ (M+N)\tilde{\delta}^{-\frac{M+N-1}{M+N}}\right ].
\end{eqnarray}

\subsubsection{The two-point function of regularized twist operators}

Let us also calculate the contribution to the Liouville action from the regularized two-point function, 
\begin{eqnarray}
S_{L,2}&=&S_{L_{1}}[\sigma_{M+N}^{\tilde{\delta}}\sigma_{M+N}^{\epsilon}]
+S_{L_{2}}[\sigma_{M+N}^{\tilde{\delta}}\sigma_{M+N}^{\epsilon}]
\end{eqnarray}

We use the map
\begin{equation}
\begin{split}
z=t^{M+N},\,\,\,\,\,\,\,\,\, t=z^{\frac{1}{M+N}},\,\,\,\,\,\,\,\, \frac{dz}{dt}=(M+N)t^{M+N-1}\\
\phi\approx \log\left [ (M+N)t^{M+N-1} \right ] +c.c.\,\,\,\,\,\,\,\,\,\, \partial_{t}\phi=\frac{M+N-1}{t}
\end{split}
\end{equation}
We have branch points at $z=0,t=0$ and $z=\infty,t=\infty$. The contribution to the Liouville action from each of these points is calculated in the same manner as the contribution from the four-point function.  We thus find:
\begin{eqnarray}
S_{L_1,2}(t=0,z=0)&=&-\frac{1}{2}(M+N-1)\log\left [ (M+N)\epsilon^{\frac{M+N-1}{M+N}}\right ]\nn
S_{L_2,2}(t=\infty,z=\infty)&=&\frac{1}{2}(M+N-1)\log\left [(M+N)\tilde{\delta}^{\frac{1-M-N}{M+N}} \right ]
\end{eqnarray}

Combining the Liouville terms for both branch points of the two-point function then gives
\begin{equation}\label{Liouville2}
S_{L,2}=\frac{M+N-1}{2}\log \left [ \tilde{\delta}^{\frac{1-M-N}{M+N}} \right ] - \frac{M+N-1}{2}\log \left [ \epsilon^{\frac{M+N-1}{M+N}} \right ].
\end{equation}

\subsubsection{Normalization of the twist operators}

In \cite{\lmone}, it was shown that the two-point function of regularized twist operators at finite separation is given by
\begin{eqnarray}
\langle \sigma_{n}^{\epsilon}(0)\sigma_{n}^{\epsilon}(a)\rangle=a^{-4\Delta_{n}} \left(n^2\epsilon^{A_{n}}Q^{B_{n}}\right)
\end{eqnarray}
where 
\begin{eqnarray}
\Delta_{n}&=&\frac{c}{24}\left(n-\frac{1}{n}\right) \,, \quad
A_{n}~=~-\frac{(n-1)^{2}}{n}\,,\quad B_{n}~=~1-n
\label{norm_terms}
\end{eqnarray}
and where $Q$ is a quantity that is regularization-dependent and which  cancels out.

The contribution from normalization term is then
\begin{equation}
\sqrt{
\frac{\langle\sigma_{M+N}^{\epsilon}(0)\sigma_{M+N}^{\epsilon}(1)\rangle}
{\langle\sigma_{2}^{\epsilon}(0)\sigma_{2}^{\epsilon}(1)\rangle
\langle\sigma_{N}^{\epsilon}(0)\sigma_{N}^{\epsilon}(1)\rangle
\langle\sigma_{M}^{\epsilon}(0)\sigma_{M}^{\epsilon}(1)\rangle}
}
=\sqrt{\frac{(M+N)^{2}\epsilon^{-\frac{(M+N-1)^{2}}{(M+N)}}}{(2MN)^{2}\epsilon^{-\frac{(N-1)^{2}}{N}-\frac{(M-1)^{2}}{M}-\frac{1}{2}}}}
\label{norm_terms-2}
\end{equation}
Combining (\ref{Liouville4}), (\ref{Liouville2}) and (\ref{norm_terms-2}), one can check that the regularization terms cancel, and so \eq{eq:4pt-2} becomes
\begin{eqnarray}\label{LiouvilleNormalization}
&&
\langle\sigma_{M+N}(\infty)\sigma_{2}(z_{0})\sigma_{N}(0)\sigma_{M}(0)\rangle 
\cr
&& {} \qquad =~
2^{-\frac{5}{4}}|a|^{-\frac{3}{4}\left(M+N\right)+\frac{1}{2}\left(\frac{M}{N}+\frac{N}{M}+1\right)}M^{-\frac{3}{4}M-\frac{1}{4}}N^{-\frac{3}{4}N-\frac{1}{4}}(M+N)^{\frac{3}{4}(M+N)-\frac{1}{4}}\,. \qquad
\end{eqnarray}

\subsection{Normalization factors for spin field insertions}

We next turn to the contributions to our amplitude (\ref{eq:main}) coming from the spin fields, 
\begin{eqnarray}
\lim_{|z|\to\infty}
\frac{\langle \mathcal{S}_4 (t(z),t_0) \rangle}{\langle \mathcal{S}_2 (t(z))\rangle} \quad
\label{eq:main-2}
\end{eqnarray}
where $\langle \mathcal{S}_4 (t(z),t_0) \rangle$ and $\langle \mathcal{S}_2 (t(z))\rangle$ are  $t$ plane  correlators (four-point and two-point respectively) of appropriately normalized spin fields, which we now define.

The normalization coefficient for a spin field $S_n^{\pm}(z_*)$ depends on the local form of the cover map at the insertion point,
\bea
\left(z-z_{*}\right)\approx b_* \left(t-{t_{*}}\right)^{n} \,.
\eea
The corresponding spin field insertion in the $t$ plane is  given by \cite{\lmtwo}
\begin{eqnarray}
b_{*}^{-\frac{1}{4n}}S^{\pm}(t_*) \,.
\end{eqnarray} 
We shall first collect all the normalization terms $b_{*}$, and we shall compute the spin field correlator in the next subsection. 

For the two-point function in the denominator, we use the map
\begin{eqnarray}
z=t^{M+N}
\end{eqnarray}
and so all the normalization factors are trivial. 

For the four-point function, taking the appropriate limits of our map
\begin{eqnarray}
z=t^{M}(t-a)^{N}
\end{eqnarray}
we find the normalization coefficients $b_{*}$ to be
\begin{equation}
\begin{split}
&z=0, t=0;\quad  z \approx (-a)^{N}t^{M} \rightarrow\quad b_{0}=(-a)^{N}\rightarrow\quad b_{0}^{-\frac{1}{4M}}=(-a)^{-\frac{N}{4M}}\\
&z=0,t=a;\quad  z\approx a^{M}(t-a)^{N}\rightarrow\quad b_{a}=a^{M}\rightarrow\quad b_{a}^{-\frac{1}{4N}}=a^{-\frac{M}{4N}}\\
&z=z_{0}, t=t_{0}=\frac{aM}{M+N};\quad z-z_{0}\approx b_{t_0} (t-t_{0})^{2}\rightarrow\quad b_{t_{0}}^{-\frac{1}{8}}\\
&z=\infty, t=\infty;\quad  z\approx t^{M+N}\rightarrow\quad b_{\infty}=1\rightarrow\quad b_{\infty}^{-\frac{1}{4(M+N)}}=1
\label{spin_norm_coeff}
\end{split}
\end{equation}
where $b_{t_0}$ was given in (\ref{bt0}).  We can therefore write the normalized spin field correlator as (writing only holomorphic parts ) 
\begin{eqnarray}
\lim_{|z|\to\infty}
\frac{\langle \mathcal{S}_4 (t(z),t_0) \rangle}{\langle \mathcal{S}_2 (t(z))\rangle}
&=&\left(b_{\infty}^{-\frac{1}{4(M+N)}}b_{t_{0}}^{-\frac{1}{8}}b_{a}^{-\frac{1}{4N}}b_{0}^{-\frac{1}{4M}}\right)\left(\frac{\langle S^{+}(\infty)S^{+}(t_{0})S^{-}(a)S^{-}(0)\rangle}{\langle S^{+}(\infty)S^{-}(0)\rangle}\right). \qquad \quad
\label{total_spin_corr}
\end{eqnarray}
Combining the holomorphic and antiholomorphic parts, the product of all spin field normalization factors is
\begin{eqnarray} \label{SpinFieldNormalization}
&&|b_{\infty}|^{-\frac{1}{2(M+N)}}|b_{t_{0}}|^{-\frac{1}{4}}|b_{a}|^{-\frac{1}{2N}}|b_{0}|^{-\frac{1}{2M}}=\cr
&&\qquad 2^{\frac{1}{4}}|a|^{-\frac{1}{2}\left(\frac{M}{N}+\frac{N}{M}+\frac{M}{2}+\frac{N}{2}-1\right)}M^{-\frac{1}{4}(M-1)}N^{-\frac{1}{4}(N-1)}(M+N)^{\frac{1}{4}(M+N-3)}\,.
\end{eqnarray}

\subsection{Spin field correlator}
Here we show the calculation of the spin field correlator term, using two different methods: firstly via bosonization  and secondly via spectral flow.
\subsubsection{Method 1: Bosonization}
Here we shall use bosonization to calculate the spin field correlators given in (\ref{total_spin_corr}). We define the bosonized fermion fields as
\begin{eqnarray}
\psi_{1}=e^{i\phi_{5}},\qquad \psi_{2}=e^{i\phi_{6}}
\end{eqnarray}
We can therefore write the spin fields as
\begin{eqnarray}
S^{\pm}(z)=e^{\pm \frac{i}{2}e_{a}\Phi^{a}(z)}
\label{spin_fields}
\end{eqnarray}
where 
\begin{eqnarray}
e_{a}\Phi^{a}(z)=\phi_{5}(z)-\phi_{6}(z) \,.
\end{eqnarray}
The OPE is
\begin{equation}
e^{i\alpha\phi(z)}e^{i\beta\phi(w)}\sim e^{i(\alpha\phi(z)+\beta\phi(w))}(z-w)^{\alpha\beta}\,.
\end{equation}
Thus we have
\begin{eqnarray}
&&\frac{\langle S^{+}(\infty)S^{+}(t_{0})S^{-}(a)S^{-}(0)\rangle}{\langle S^{+}(\infty)S^{-}(0)\rangle}=\lim_{t\to\infty}\frac{\langle S^{+}(t)S^{+}(t_{0})S^{-}(a)S^{-}(0)\rangle}{\langle S^{+}(t)S^{-}(0)\rangle}\cr
&&\quad=\lim_{t\rightarrow \infty}
\frac{
\langle :\!\exp\left(\frac{i}{2}e_{a}\Phi^{a}(t)\right)\!\!:\,:\!\exp\left(\frac{i}{2}e_{a}\Phi^{a}(t_{0})\right)\!\!:\,:\!\exp\left(-\frac{i}{2}e_{a}\Phi^{a}(a)\right)\!\!:\,:\!\exp\left(-\frac{i}{2}e_{a}\Phi^{a}(0)\right)\!\!:\rangle
}{
\langle :\! \exp\left(\frac{i}{2}e_{a}\Phi^{a}(t)\right)\!\!:\,:\!\exp\left(-\frac{i}{2}e_{a}\Phi^{a}(0)\right)\!\!:\rangle 
}
\,.\cr
&&
\label{spin_corr}
\end{eqnarray}

For the four-point correlator, we obtain
\bea
&&\langle :\!\exp\left(\frac{i}{2}e_{a}\Phi^{a}(t)\right)\!\!:\,:\!\exp\left(\frac{i}{2}e_{a}\Phi^{a}(t_{0})\right)\!\!:\,:\!\exp\left(-\frac{i}{2}e_{a}\Phi^{a}(a)\right)\!\!:\,:\!\exp\left(-\frac{i}{2}e_{a}\Phi^{a}(0)\right)\!\!:\rangle \cr
&&{}\qquad =~(t-t_{0})^{\frac{1}{2}}(t-a)^{-\frac{1}{2}}(t_{0}-a)^{-\frac{1}{2}}t^{-\frac{1}{2}}t_{0}^{-\frac{1}{2}}a^{\frac{1}{2}}\,.
\label{4pt_spin_contr}
\eea
For the two-point correlator, we obtain
\bea
\langle :\! \exp\left(\frac{i}{2}e_{a}\Phi^{a}(t)\right)\!\!:\,:\!\exp\left(-\frac{i}{2}e_{a}\Phi^{a}(0)\right)\!\!:\rangle &=& t^{-\frac12} \,.
\label{2pt_spin_contr}
\eea
Combining (\ref{4pt_spin_contr}) and (\ref{2pt_spin_contr}), we find that (\ref{spin_corr}) is given by
\begin{eqnarray}
\frac{\langle S^{+}(\infty)S^{+}(t_{0})S^{-}(a)S^{-}(0)\rangle}{\langle S^{+}(\infty)S^{-}(0)\rangle}&=&\lim_{t\to\infty}(t-t_{0})^{\frac{1}{2}}(t-a)^{-\frac{1}{2}}(t_{0}-a)^{-\frac{1}{2}}t_{0}^{-\frac{1}{2}}a^{\frac{1}{2}}\cr
&=&(t_{0}-a)^{-\frac{1}{2}}t_{0}^{-\frac{1}{2}}a^{\frac{1}{2}}\cr
&=&(-a)^{-\frac{1}{2}}N^{-\frac{1}{2}}M^{-\frac{1}{2}}(M+N)\,.
\label{spin_contr_calc}
\end{eqnarray}
Combining (\ref{spin_contr_calc}) with the antiholomorphic part, we obtain
\bea
\frac{\langle S^{+}(\infty)S^{+}(t_{0})S^{-}(a)S^{-}(0)\rangle}{\langle S^{+}(\infty)S^{-}(0)\rangle}=\frac{(M+N)^2}{MN} \frac{1}{|a|}\,.
\label{eq:spinfieldans2}
\end{eqnarray}

\subsubsection{Method 2: Spectral Flow}

We now compute the spin field correlator using the sequence of spectral flow transformations and coordinate changes described in Section \ref{sec:sf}. This serves as a cross-check of the above calculation via bosonization.

Having lifted to the $t$ plane, and taken care of the normalization factors of the spin field insertions, we are left with the $t$ plane spin field correlator (writing holomorphic fields only)
\begin{eqnarray}
&&{}_t \bra{0_{R,-}} S^- (a) S^+ \left(\tfrac{M}{M+N}a\right)\ket{0_R^-}_t \,.
\end{eqnarray}
We recall that the action of spectral flow is straightforward for operators where the fermion content may be expressed as a simple exponential in the language in which the fermions are bosonized. For such operators with charge $j$, spectral flow with parameter $\alpha$ gives rise to the transformation
$$\hat O_j (t)\to t^{-\alpha j}\hat O_j (t).$$
We first spectral flow by $\alpha = 1$ in the $t$ plane. 
This gives
\begin{eqnarray}
\sqrt{\frac{M+N}{M}}~
{}_t \bra{0_{NS}} S^- (a) S^+ \left(\tfrac{M}{M+N}a\right)\ket{0_{NS}}_t.
\end{eqnarray}
Next, we change coordinate to
\begin{eqnarray}
t'& =& t - \tfrac{M}{M+N}a \,.
\end{eqnarray}
This gives the $t'$ plane correlator
\begin{eqnarray}
\sqrt{\frac{M+N}{M}}~
{}_{t'} \bra{0_{NS}} S^- \left(\tfrac{N}{M+N} a\right) \ket{0_R^+}_{t'} \;.
\end{eqnarray}
We then spectral flow by $\alpha = -1$ in the $t'$ plane, yielding
\begin{eqnarray}
&&\sqrt{\frac{M+N}{M}} \left(\frac{N}{M+N} a \right)^{-\frac{1}{2}}
{}_{t'} \bra{0_{R, -}} S^- \left(\tfrac{N}{M+N} a\right) \ket{0_{NS}}_{t'} \\
&=&\left(\frac{M+N}{\sqrt{MN}} a^{-\frac{1}{2}}\right) \,
{}_{t'} \bra{0_{R,-}} S^- \left(\tfrac{N}{M+N} a\right) \ket{0_{NS}}_{t'}.
\end{eqnarray}
Next, we change the coordinate to
\begin{eqnarray}
&&\hat{t}=t'-\tfrac{N}{M+N}a \,.
\end{eqnarray}
This gives
\begin{eqnarray}
\left(\frac{M+N}{\sqrt{MN}} a^{-\frac{1}{2}} \right)
{}_{\hat{t}} \braket{0_{R,-}}{0_{R}^{-}}_{\hat{t}} &=& \frac{M+N}{\sqrt{MN}} a^{-\frac{1}{2}} \,.
\end{eqnarray}
The final spectral flow by $\alpha =1$ in the $\hat t$ plane has no effect.

Adding in the anti-holomorphic factors, we obtain
\bea\label{SpinFieldCorrelator}
\frac{(M+N)^2}{MN} \frac{1}{|a|},
\eea
in agreement with \eq{eq:spinfieldans2}.

\bigskip

Combining the Jacobian factor (\ref{Jacobian}), the Liouville action contribution (\ref{LiouvilleNormalization}), the spin field normalization factors (\ref{SpinFieldNormalization}), and the spin field correlator (\ref{SpinFieldCorrelator}), we find
\begin{equation}
C_{MN}=\frac{M+N}{2MN} \,.
\end{equation}

\end{appendix}

\newpage

\begin{adjustwidth}{-.5mm}{-0.5mm} 


\providecommand{\href}[2]{#2}\begingroup\raggedright\endgroup

\end{adjustwidth}

\end{document}